\documentclass{jfm}
\usepackage{graphicx}
\usepackage{subcaption}
\usepackage{epstopdf, epsfig}
\usepackage{xcolor}
\usepackage{natbib}
\usepackage{mathtools}
\usepackage{tikz}
\usepackage{bm}
\usepackage{sansmath}
\usepackage{amsmath}
\usepackage{lineno}
\usepackage[hypertexnames=false]{hyperref}
\usepackage[export]{adjustbox}

\usepackage[linesnumbered, ruled, vlined]{algorithm2e}
\SetKw{Break}{break}

\newcommand{\We}{W\kern-0.15em e\,}
\newcommand{\Ohd}{O\kern-0.04em h_d\,}
\newcommand{\Bo}{B\kern-0.04em o\,}
\newcommand{\Dr}{D\kern-0.04em r\,}
\newcommand{\Wed}{W\kern-0.15em e _d\,}
\newcommand{\LapH}[1]{\frac{ #1^{k+1}_{i-1}-2  #1^{k+1}_i+ #1^{k+1}_{i+1}}{\delta_r^2} + 
    \frac{#1^{k+1}_{i+1} - #1^{k+1}_{i-1}}{2 (i-1) \delta_r^2}}
\newcommand{\bfjfm}[1]{\bm{\mathsfbi{#1}}}

\newcommand{\mycomment}[1]{}

\shorttitle{Droplet rebounds on a fluid interface}
\shortauthor{E. Ag\"uero and others}

\title{Droplet rebounds off a fluid bath at low Weber numbers}

\author{Elvis A. Ag\"uero\aff{1,2}\corresp{\email{elvisaguero@brown.edu}},
  Carlos A. Galeano-Rios\aff{3,4,5}
  \corresp{\email{grcarlosa@gmail.com}},
  Clodoaldo Ragazzo\aff{6},
  Chase T. Gabbard\aff{1},  
  Daniel M. Harris\aff{1}
 \and Paul A. Milewski\aff{7,4}}

\affiliation{
\aff{1}School of Engineering, Brown University, Providence, RI 02912, USA
\aff{2}ILACVN, Universidade Federal da Integração Latino-Americana, Foz de Iguaçu, 85867-970 PR, Brazil
\aff{3}College of Computational Sciences, Minerva University, San Francisco, CA 94103, USA
\aff{4}Department of Mathematical Sciences, University of Bath, Bath, BA2 7AY, UK
\aff{5}School of Mathematics, University of Birmingham,  Birmingham, B15 2TT, UK
\aff{6}Instituto de Matemática e Estatística, Universidade de São Paulo, São Paulo, SP 05508-090, Brazil
\aff{7}Department of Mathematics, The Pennsylvania State University, University Park, State College, PA 16802, USA}

\begin{document}

\maketitle

\begin{abstract}
We present a method to simulate non-coalescing impacts and rebounds of droplets onto the free surface of a liquid bath, together with new experimental data, focused on the low-speed impact of droplets. The method is derived from first principles and imposes only natural geometric and kinematic constraints on the motion of the impacting interfaces, yielding predictions for the evolution of the contact area, pressure distribution, and wave field generated on both impacting masses. This work generalises an existing kinematic-match method whose prior applications dealt with deformation of the surface of the bath only; i.e., neglecting that of the droplet. The method’s extension to include droplet deformation gives predictions that compare favourably with existing experimental results and our new experiments conducted in the low-Weber-number regime.\end{abstract}


\section{Introduction}
Droplet impacts are integral to a diverse range of natural and industrial processes \citep{MohammadKarim2023review,hu2024understanding}. For example, raindrop impacts on liquid and compliant biological surfaces facilitate pathogen transport \citep{impact_xu_2024,coherent_wu_2024, Shen_Kulkarni_Jamin_Popinet_Zaleski_Bourouiba_2025}.
Here, we present new insights from experiments and modeling of low-Weber-number impacts of submillimetric droplets on liquid baths, a regime governing the terminal bouncing and coalescence of agricultural and biological sprays.

Since the seminal work of \citet{Worthington1882}, researchers have sought to understand droplet impacts on a wide range of surfaces \citep{Josserand2016-2AnnualReviews,REIN199361}. The majority of these efforts have focused on moderate and high Weber number impacts, which can result in spreading, retraction, or splashing \citep{Yarin2006} for impacts on solid surfaces, and also corona splashing and coalescence on liquid surfaces \citep{highspeed_thoroddsen_2008,kavehpour2015,Burzynski_Roisman_Bansmer_2020}. Notably, whether a droplet coalesces with a bath during impact depends upon whether the thin gas layer separating the liquid bodies can be sufficiently drained during their interaction  \citep{CouderEtAl2005Noncoalescence,Yarin2006,kavehpour2015}. Thus, at high Weber numbers, coalescence is observed readily due to rapid gas layer drainage, while at moderate Weber numbers, impacting droplets partially or fully rebound \citep{AlventosaEtAl2023,Burzynski_Roisman_Bansmer_2020}, provided that the separating gas film does not thin below a critical thickness \citep{sprittles2024dropbounces, SprittlesImpact2020}. Interestingly, while reducing the bath height leads to substrate-independent rebound properties \citep{sanjay_jfm_2023,GabbardEtAl2025}, vertically oscillating  the bath can replenish the intervening gas layer and sustain droplet bouncing \citep{CouderEtAl2005Noncoalescence}. 

Droplets bouncing on vibrating liquid baths have attracted significant interest due to their dynamics showing quantum-like behaviour reminiscent of pilot-wave theory \citep{BushAndOza2020}. More recently, droplets bouncing on baths driven at two superimposed frequencies have been shown to self-propel through interaction with their own wavefields, giving rise to collective behaviours and the emergence of `superwalkers’, droplets larger than the maximum stable size on single-frequency baths \citep{ValaniEtAl2019}. \citet{Galeano-RiosEtAl2019} derived a model for superwalkers with detailed features for the wave field. Their work treated the bouncing droplet as a rigid sphere and only partially captured the dynamics of superwalkers, suggesting that droplet deformation plays an integral role. 

For droplet impact on non-vibrating baths, whether a droplet bounces or coalesces depends on the fluid properties, drop size, impact speed, and bath depth \citep{CouderEtAl2005Noncoalescence,Yarin2006,kavehpour2015,tang2019bouncing}. For fixed droplet size and fluid properties, the bouncing regime is bounded by two transitions: a low-speed transition from coalescence to bouncing, and a high-speed transition from bouncing back to coalescence \citep{pan2007dynamics,zhao2011transition}. 
At moderate Weber number, the contact time between the droplet and bath, and the coefficient of restitution, which quantifies the translational energy recovery, are approximately constant, but become dependent on impact speed at low Weber numbers \citep{jayaratne1964coalescence,zhao2011transition,zou2011experimental,wu2020small,Sanjay_Chantelot_Lohse_2023}, consistent with trends observed for droplet impact on non-wetting rigid substrates \citep{GabbardEtAl2025}. Although bath-depth effects can alter rebound metrics \citep{pan2007dynamics,zou2011experimental} and the early-time dynamics of the air film have been shown to vary with impact conditions \citep{highspeed_thoroddsen_2008}, we focus here on impacts on deep baths, which have been shown to yield robust rebound properties \citep{AlventosaEtAl2023}. While these studies capture key rebound scalings at moderate Weber numbers, low Weber number impacts, and models capable of resolving the droplet and bath deformation remain comparatively sparse. 

Fully resolving the dynamics of a deformable impactor is challenging, as numerical multiscale effects typically span three orders of magnitude \citep{CouderEtAl2005Noncoalescence}, and natural time scales are short \citep[see figure 3b of][]{GaleanoRiosEtAl2017}. Early attempts to address these difficulties relied on asymptotic analysis and linearised formulations of the free surface equations of motion for impacts against inviscid, incompressible fluids \citep{Wagner1932, korobkin2004analytical,HowisonEtAl1991}. A weakly viscous framework introduced by \citet{Lamb}, and revisited by \citet{DiasEtAl2008}, was combined with a formulation accounting for drop-bath coupling through the \emph{kinematic match} (KM) model \citep{GaleanoRiosEtAl2017,Galeano-RiosEtAl2019,Galeano-RiosEtAl2021}. This approach imposes natural geometric and kinematic constraints to solve for the evolution of the contact area between the impacting bodies. The KM model does not take into account the thin gas film, replacing it with an infinitely thin contact surface that transmits the pressure between the fluid bodies. Although providing remarkable agreement with experiments in the weakly viscous, low-to-moderate Weber number regime, existing implementations have been restricted to rigid impactors. 

Recently, \citet{AlventosaEtAl2023} introduced a simplified version of the KM framework that allowed both impacting bodies to deform through orthogonal mode decompositions, enforces contact at a single point, and assumes a polynomial pressure profile at contact. This \textit{1-point kinematic match} (1PKM) model, predicts droplet rebound dynamics near the inertio-capillary regime, including trajectories, coefficient of restitution, contact times, and maximum deflection, in close agreement with experiments and direct numerical simulations (DNS). \citet{phillips2024rebound2D} and \citet{phillips2024lubrication} solved directly for the pressure at contact by accounting for the gas film mediating rebound between the droplet and bath. These models are particularly well-suited for weakly viscous flows impacted at low speeds, and were validated against direct numerical simulations and rigid-body impacts. Other reduced-order models have incorporated droplet deformation by representing the droplet as a spring–mass system \citep{Blanchette2016,Blanchette2017}, initially using a single vertical spring coupled to a linear bath response and later an octahedral spring network calibrated to the drop’s natural oscillation modes. 

In this manuscript, we develop a KM model from first principles that allows both the droplet and the bath surface to deform upon impact. Previous KM-based approaches treated the droplet as approximately rigid \citep{Galeano-RiosEtAl2019} and, while successful in predicting integral rebound quantities, such as the coefficient of restitution, contact time, and surface pressure profiles, remained limited when droplet deformation became non-negligible. Building on the weakly viscous bath formulation of \citet{Galeano-RiosEtAl2021} and \citet{AlventosaEtAl2023}, the present model imposes full kinematic match conditions, allowing droplet deformation through a spectral representation of the surface of the droplet in spherical coordinates using Legendre polynomials, as done by \cite{phillips2024lubrication} and \cite{AlventosaEtAl2023}. This new framework yields detailed predictions for the evolution of the pressure field and pressed area without explicitly resolving the intervening gas layer. We also extend experimental measurements to submillimetric drops impacting at velocities approaching the lower rebound limit, reporting contact time and coefficient of restitution. These measurements fill a previously sparse region of parameter space and provide stringent benchmarks for models of droplet–bath impacts.

The manuscript is organized as follows. Section \ref{sec:Problem_formulation} states the governing equations, the simplifications and the boundary conditions. Section \ref{sec:discretised_model} presents a detailed description of the discretised system of equations to be solved numerically, complemented by the details of the numerical implementation in appendices \ref{section:Appendix_L} and \ref{sec:comp_implementation}.  Section \ref{sec:results} presents the main results and comparisons with previous modelling efforts, including experimental results from the literature \citep{AlventosaEtAl2023}, and new experiments performed in the low-$\We$ regime. Finally, section \ref{sec:Discussion} discusses our main results and contextualizes them using available literature. 
\section{Experimental set-up}\label{app:experiment}

\begin{figure}
    \centering \includegraphics[width=\linewidth]{Images/figure1.pdf}
    \caption{($a$) A rendering of the experimental setup. ($b$) Height of the centre of mass $h$ against time $t$ for a
bouncing drop, where $\Wed=1.285$, $\Ohd=0.0238$, and $\Bo=0.049$. The dashed lines are parabolic fits to the incoming and outgoing data, and the dotted line denotes the drop radius $R_{d} = 0.328$ mm. ($c$) Select experimental images corresponding to panel ($b$). The dashed line indicates the height of the undisturbed free surface at the point of contact.}
    \label{fig:experiment}
\end{figure}

We performed experiments using submillimetre-sized silicone oil droplets. The droplets were produced by a piezoelectric droplet generator mounted on a linear stage, allowing precise control over droplet size and release \citep{harris2015low,ionkin2018note}, as shown in Figure~\ref{fig:experiment}($a$). Details on the generator design and droplet formation process are available in previous work \citep{GabbardEtAl2025}. The radius of the droplet $R_{d}$ and its impact velocity $V_{0}$ were controlled by adjusting the nozzle size and its initial height above the bath. Both the droplet and bath were the same liquid: silicone oil with density $\rho = \rho_{d}= 870$ kg m$^{-3}$, dynamic viscosity $\mu =\mu_{d} = 0.00174$ g cm$^{-1}$ s$^{-1}$, and surface tension $\sigma = \sigma_{d} = 18.7$ g s$^{-2}$. All experiments were performed in air at room temperature.

The experiment begins by filling a clear, rectangular container (68 mm $\times$ 68 mm $\times$ 37 mm) with silicone oil slightly beyond brimful conditions, ensuring unobstructed side-view imaging of the droplet’s contact with the bath. The connecting tubing is purged to eliminate trapped air \citep{harris2015low}, and the pulse width of the piezoelectric disk is tuned to produce an upward-moving droplet, allowing interfacial oscillations to subside before impact \citep{GabbardEtAl2025}. For the lowest $\We$ experiments, a subsequent droplet impact was required. In these cases, both droplet and bath deformations were observed to decay before the next impact, and any effect of remnant deformations on rebound metrics was smaller than the experimental uncertainty. The nozzle is then positioned $\approx$1 mm above the free surface to release a droplet, which falls under gravity and impacts the bath, resulting in rebound, coalescence, or transient floating. Herein, we focus on rebounds, which are defined as a rebounding droplet whose south pole rises above the undisturbed interface. For each nozzle height, three to six trials are conducted before systematically adjusting the height to achieve the desired impact velocity. This process is repeated until rebound ceases, which occurs at both sufficiently low and high impact velocities.

The droplet silhouette was captured using a Phantom Miro LC311 high-speed camera with a Laowa 25 mm ultra-macro lens. Most experiments were recorded at 15,000 fps, except for those with the smallest droplets, which required a higher frame rate of 39,000 fps due to their shorter contact times. Spatial calibration was performed using a microscope calibration slide, yielding an average resolution of 0.00469 mm pixel$^{-1}$. Droplet shape and trajectory were extracted from the recordings using MATLAB. The droplet's centre of mass was calculated in each frame by assuming a surface of revolution about the vertical axis. The droplet height $h$ was defined as the vertical distance from the centre of mass to the undisturbed bath surface. Figure~\ref{fig:experiment}($b$) plots $h$ versus time $t$ for a representative experiment, with blue and gold markers showing the incoming and outgoing trajectories, respectively. Time $t=0$ marks the first visible contact between the droplet and bath, and the gray region indicates the contact period during which $h$ cannot be reliably determined. The droplet radius $R_{d}$ is determined using the projected area of the droplet in the frames prior to contact and assuming the projected shape is a circle. The drop velocity at impact $V_{0} = V\left(t=0\right)$ was determined by fitting a parabola to  $h\left(t\right)$ for $t<0$ and evaluating its slope at $t=0$. The rebound velocity $V_{r}=V\left(t=t_{c}\right)$ was computed by fitting a parabolic fit to $h\left(t\right)$ for $t>t_{c}$ and evaluating its slope at $t=t_{c}$, where $t_{c}$ is the contact time. The uncertainty of each measurement was determined using the spatial and temporal resolution of our experimental set-up, corresponding to the width of a pixel and time between successive frames, respectively. Uncertainty propagation was performed using Mathematica, where first-order Taylor expansion was used for uncertainty propagation and uncertainties were assumed uncorrelated.

Figure~\ref{fig:experiment}($c$) shows experimental images at select moments: during free fall ($t=-1.4$ ms), initial contact with the bath ($t=0$ ms), maximum vertical compression during impact ($t=1.3$ ms), liftoff from the surface ($t=6.8$ ms), and during rebound ($t=8.9$ ms). Liftoff is defined as the first frame where the droplet’s south pole rises above the undisturbed bath surface. Due to post-impact oscillations, the droplet is typically non-spherical at liftoff. Therefore, in addition to measuring the rebound velocity $V_{r}$, we also extract the height of the centre of mass of the drop at rebound $h_{r}=h\left(t=t_{c}\right)$. 

\section{Problem formulation}\label{sec:Problem_formulation}

 We consider the three-dimensional incompressible flow of a fluid bath of infinite depth and infinite lateral extension, which is initially at rest. The fluid has density $\rho$, kinematic viscosity $\nu$, and surface tension $\sigma$. At time $t=0$, the undisturbed free surface of the bath is subject to a non-coalescing normal impact by a droplet of density $\rho_d$, kinematic viscosity $\nu_d$ and surface tension $\sigma_d$. The interface of the droplet is assumed to be spherical at the moment of first contact, having a radius $R_d$ and zero relative surface velocity with respect to the centre of mass of the droplet. 
 
\begin{figure}
    
\centering
\includegraphics[width = .60\textwidth]{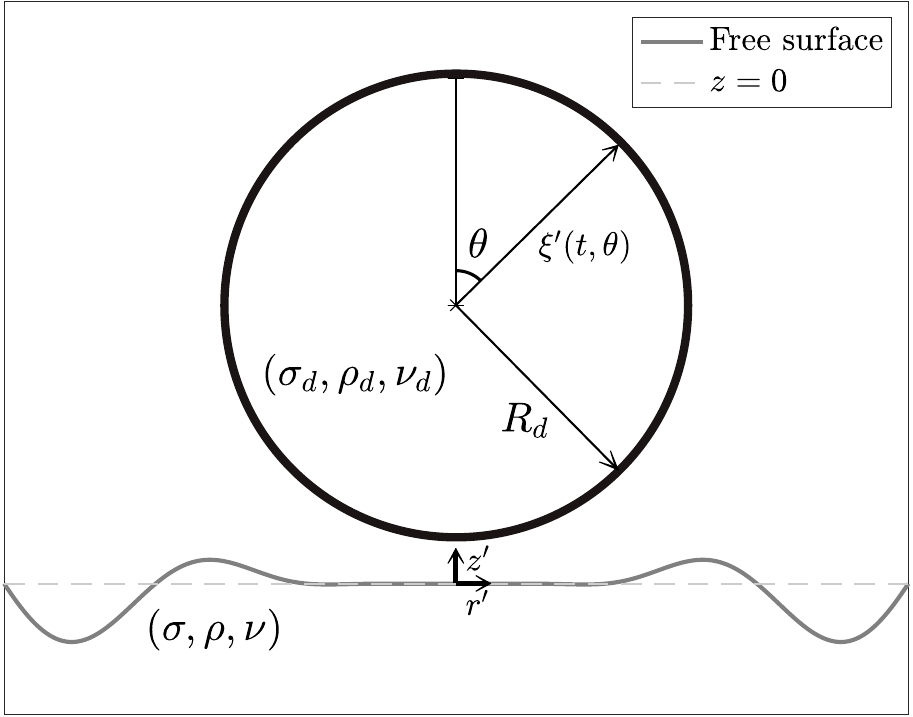}
\caption{Schematic of the problem. An undeformed droplet (thick solid black line)
of radius $R_d$ and fluid properties $(\sigma_d, \rho_d, \nu_d)$ is depicted
above the surface of a fluid bath with properties $(\sigma, \rho, \nu)$, shown with a solid grey line. The surface of the droplet is described in an non-inertial spherical reference frame by $\xi'(t, \theta)$, whose origin is the centre of mass of the droplet. The height of the bath's surface, denoted by $\eta(r, t)$, is described in a fixed cylindrical reference frame whose origin coincides with the initial point of impact. }
 \label{fig:schematic_before_imapct}
\end{figure}

 We use cylindrical coordinates $(r,\varphi,z,t)$ with the $z$ pointing normally to the undisturbed free surface and away from the bath (i.e., gravity is given by $-g\hat{\bf{z}}$), with $z = 0$ corresponding to the undisturbed free surface, as illustrated in figure 
\ref{fig:schematic_before_imapct}. The axial symmetry of the problem around the $z$ axis allows us to ignore the dependence on $\varphi$, as is henceforth done.

 \subsection{The bath model}
 We model the bath using a linearised, quasi-potential flow approximation, initially presented by \cite{Lamb} and revisited by \cite{DiasEtAl2008}, and follow closely the notation and formulation presented by \cite{GaleanoRiosEtAl2017}. This fluid model assumes that the free-surface elevation can be described by $z = \eta(r, t)$, and introduces a velocity potential $\phi(r, z, t)$, as well as the pressure on the free surface due to the effect of the impacting droplet $p(r, t)$. We further assume that the integrity of the separating air layer is maintained throughout the impacts and simplify our model by assuming that this air layer is infinitely thin and serves only to transfer pressure between the drop and the bath. This treatment is justified in the low-Weber-number regime as a limiting case of lubrication theory where the pressure variation across the film is negligible to leading order, an assumption supported by the stable rebounds observed in our experiments. Furthermore, the impact is assumed to be perfectly non-wetting, so that the contact angle where the free surface detaches from the drop is constant and equal to $180^\circ$ at all times during contact.
 
 We define $V_0$ to be the impacting velocity of the droplet with the bath at $t=0$, when contact is initiated (i.e., the centre of mass of the droplet is at a distance $R_d$ above the free surface). 

Taking $R_d$, $T_d = \sqrt{\rho_d R_d^3 / \sigma_d}$ and $\rho_d R_d^3$ to be the characteristic length, time and mass, respectively, we adopt the following dimensionless numbers:
\begin{equation}
    \label{eqn:adim_parameters}
    \Rey = \frac{R_d^2}{T_d \nu}, \quad  Bo = \frac{gT_d^2}{R_d}, \quad \Ohd = \frac{\nu_d T_d}{R_d^2},  \quad \Wed = \frac{\rho_d V_0^2 R_d}{\sigma_d}, \quad \Dr = \frac{\rho_d}{\rho}, \quad Sr = \frac{\sigma_d}{\sigma}.
\end{equation}
that is, the Reynolds, Bond, Ohnesorge and Weber numbers, the density ratio (similar to an Atwood number), and the ratio of the surface tensions of the two fluids.
The non-dimensional quasi-potential formulation \citep{GaleanoRiosEtAl2017} results in the following set of equations:
\begin{equation}
    \label{eqn:km_2}
    \partial_t \eta =\frac{2}{\Rey} \Delta_H \eta + \partial_z \phi, \quad z = 0;
\end{equation}
\begin{equation}
    \label{eqn:km_3}
   \phi_t = -\Bo \eta + Sr\, Dr\, \kappa[\eta] + \frac{2}{\Rey} \Delta_H \phi  - Dr \,p , \quad z = 0;
\end{equation}
which are subject to 
\begin{equation}
    \eta, \phi, |\nabla \phi| \to 0 \  \text{as} \ r\to\infty, \ \text{for}\  z = 0,
\end{equation}
and
\begin{equation}
    \eta(r,0) = 0,\quad\forall r\geq0,
\end{equation}
\begin{equation}
    \phi(r,z,0) = 0,\quad \forall r\geq 0,\  z\leq 0.
\end{equation}
Here, $\Delta_H = \partial_{rr} + (1/r)\,\partial_r$ denotes the two-dimensional Laplacian for axially symmetric functions, and $\kappa$ is twice the mean-curvature operator; after linearization, $\kappa$ is reduced to $\Delta_H$. The function $\phi$ is harmonic on its domain ($\Delta\phi = 0,\ z \leq 0$), thus allowing for the definition of the Dirichlet-to-Neumann operator for the Laplace problem in said domain. This operator is given by:
\begin{equation}
    \label{eqn:dirichlet-to-neumann}
    \partial_z \phi(r, z = 0, t) = N \Phi(r, t) = \frac{1}{2\pi} \lim_{\epsilon\to 0^+} \int_{\mathbb{R}^2 \setminus B(r, \epsilon)} \frac{\Phi(r, t) - \Phi(s, t)}{|r-s|^3} dA(s)
\end{equation}
\citep[see][]{GaleanoRiosEtAl2017}, in which, the correspondence $\{(x, y, 0) \in \mathbb{R}^3\} \equiv \mathbb{R}^2$ has been made, and $\Phi(r,t)\coloneqq \phi(r,z=0,t)$. Therefore, equation (\ref{eqn:dirichlet-to-neumann}), together with axial symmetry, reduces this three-dimensional problem to a one-dimensional configuration. 

\subsection{The droplet model}
The droplet is subject to gravitational forces and to the pressure distribution from the air layer that separates the bath and droplet in the non-coalescing impact. The pressure on the droplet is assumed to be identical to the pressure on the bath, as would be expected in the lubrication limit when the intervening air layer is thin.
\begin{figure}
    \centering
    \includegraphics[width = .60\textwidth]{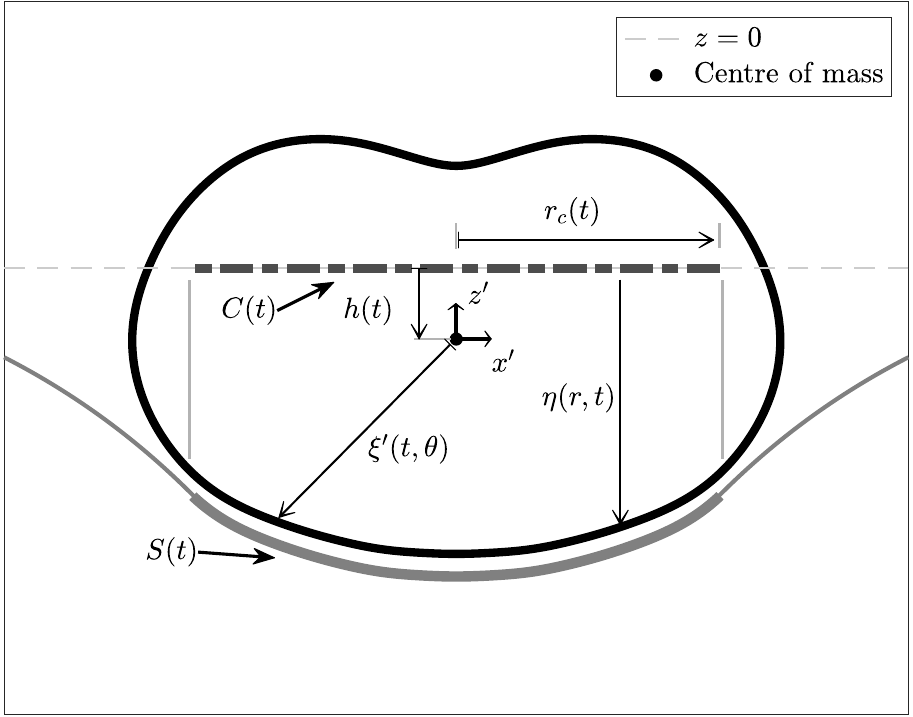}
    \caption{Schematic of deformations during impact. The surface of the bath is shown with thin grey solid lines outside the pressed surface $S(t)$, and with a thick grey solid line in the pressed surface $S(t)$. The droplet interface is shown with a thick black line. The orthogonal projection of $S(t)$ onto the $z = 0$ plane is $C(t)$, shown with a thick dark grey dashed line. Variables $h(t)$, $\eta(r, t)$ and $r_c(t)$ correspond to the height of the centre of mass of the droplet, the elevation of the free-surface of the bath and the radius of $C(t)$, respectively. The origin of the $(x',z')$ system of reference is attached to the centre of mass of the droplet. The separation between the droplet and the bath over the pressed surface is introduced solely for the purpose of better visualisation of the two, as droplet and bath interfaces are predicted to fully coincide within $S(t)$.} 
    \label{fig:schematic_impact}
\end{figure}

\subsubsection{The centre of mass of the droplet}
The non-dimensional height $h(t)$ of the centre of mass of the droplet is therefore governed by
\begin{equation}\label{eqn:vel_cm}
    \dot{h}(t) = v(t),
\end{equation}
\begin{equation}
    \label{eqn:centerofmass_drop}
    \dot{v}(t) = -Bo + \frac{3}{4\pi } \int_{C(t)} p(r,t) dA,
\end{equation}
which are subject to
\begin{equation}
    h(0) = 1,
\end{equation}
and 
\begin{equation}
    v(0) = -\sqrt{\Wed}
\end{equation}
where $C(t)$ is the orthogonal projection of the pressed surface $S(t)$ onto the $z = 0$ plane (see figure \ref{fig:schematic_impact})
and the same units as with the bath model where used for non-dimensionalisation. Moreover, we assume that $C(t)$ is simply connected; and, therefore, as follows from the axial symmetry assumptions, is a disk of radius $r_c(t)$.

\subsubsection{Droplet deformation}
To analyse the droplet deformation, we use spherical coordinates $(\xi,\theta,\tau)$ whose origin is at the centre of mass of the droplet. Figure \ref{fig:schematic_impact} shows that the spherical coordinates are based on the axes $x'$ and $z'$, which were chosen to be respectively parallel to the $r$ and $z$ axes from the cylindrical coordinates that are used to describe the fluid of the bath. That is,
\begin{equation}
    x' = r,\ \  z' = z-h(t),\ \ \xi' = \sqrt{
    r^2+\left(z'\right)^2},\ \ \theta = \arccos\left(\frac{z'}{\xi'}\right).
\end{equation}
Moreover, the axial symmetry of the problem implies that we can ignore the dependence on the azimuthal angle $\tau$.

We assume that the radial coordinate that describes the shape of the drop $\xi'$ can be written in terms of a perturbation away from a spherical shape, $\xi'(\theta,t) = R_d + \zeta(\theta,t)$. We also adopt the hypothesis of small deformations, that is, terms of the order of $\zeta^2$ are neglected, and we assume $\partial_{\theta} \zeta \ll 1$. We can thus express the surface perturbation as 
\begin{equation}
    \label{eqn:Legendre_decomposition_1}
    \zeta(\theta, t) = \sum\limits_{l = 2}^{\infty} \mathcal{A}_l(t) P_l(\cos(\theta)),
\end{equation}
where $P_l$ is the $l$th-order Legendre polynomial. These polynomials form a complete set of orthogonal functions defined on the interval $[-1, 1]$. The assumptions of small deformations and fluid incompressibility imply that the term $l =0$ does not appear in equation (\ref{eqn:Legendre_decomposition_1}). The term $l=1$ is also absent from (\ref{eqn:Legendre_decomposition_1}) because the centre of mass is fixed at $z'=0$. 

We also decompose the pressure distribution due to contact, obtaining
\begin{equation}
    p\left(\theta,t\right) = \sum_{l = 0}^{\infty} \mathcal{B}_l(t) P_l\left(\cos(\theta)\right).
\end{equation}
We note that all terms $l\geq 0$ appear in the sum in this case. We recover $B_l$ from $p(\theta, t)$ via
\begin{equation}
    \mathcal{B}_l(t) = \frac{1}{2l + 1} \int_{0}^{\pi} p(\theta, t) P_l(\cos(\theta)) \sin(\theta) d\theta.
    \label{eqn:Legendre_projection}
\end{equation}
The surface modes satisfy the following set of equations \citep{AlventosaEtAl2023, Lamb,phillips2024lubrication,Blanchette2016}: 
\begin{equation}
    \label{eqn:drop_governing}
    \dot{\mathcal{A}}_l = \mathcal{U}_l , \text{ for } \ l \geq 2,
\end{equation}
\begin{equation}
    \label{eqn:drop_governing_1}
    R_d^3 \rho_d \dot{\mathcal{U}}_l = - \sigma_d l(l+2)(l-1) \mathcal{A}_l- 2 R_d \mu_d (2l+1)(l-1)\mathcal{U}_l-R_d^2l\mathcal{B}_l , \text{ for } \ l \geq 2,
\end{equation}
where $\mu_d = \rho_d \nu_d$ is the dynamic viscosity of the droplet. Observe that, when the pressure terms are zero in (\ref{eqn:drop_governing_1}), the uncoupled surface modes of the drop are damped harmonic oscillators with natural frequencies $w_l^2 = l(l-1)(l+2) \sigma_d / \left( \rho_d R_d^3\right)$ \citep{Lamb}. This motivates the choice for characteristic time, $T_d$, in (\ref{eqn:adim_parameters}), as it has the same order of magnitude as the period of the first modes of oscillation.

Using the same characteristic units as for the bath equations, the dimensionless form of equations (\ref{eqn:drop_governing}) and (\ref{eqn:drop_governing_1}) is
\begin{equation}
    \label{eqn:drop_governing_adim}
    \dot{A}_l = U_l, \text{ for } \ l \geq 2,
\end{equation}
\begin{equation}
    \label{eqn:drop_governing_adim_1}
    \dot{U}_l = - l(l+2)(l-1) A_l-2\Ohd (2l+1)(l-1)U_l-lB_l , \text{ for } \ l \geq 2;
\end{equation}
where $A_l = \mathcal{A}_l/R_d$, $U_l=\mathcal{U}_lT_d/R_d$,  $B_l = \mathcal{B}_l R_d/ \sigma_d$, which are subject to
\begin{equation}
    A_l(0) = 0,\ U_l(0) = 0, \text{ for } \ l \geq 2.
\end{equation}

We note that equations (\ref{eqn:drop_governing})  and (\ref{eqn:drop_governing_1}) are valid in the weakly viscous limit, which requires the condition $\Ohd \sqrt{l} < 0.03$ \citep{MolacekBush2012}. We satisfy this condition by truncating the Legendre series in a way that preserves the inequality; see table \ref{tab:parameters}.

\subsection{The contact model}
We define $r_M(t)$ as the smallest distance from the $z$ axis to the point on the surface of the droplet for which the plane that is tangent to the interface of the droplet is vertical. That is, $r_M(t)$ is an upper bound for the maximum possible contact radius that a given drop shape can have when in contact with the fluid. 

We can then introduce the function $z_d(r, t)$ such that $h(t) + z_d(r, t)$ describes the vertical height of the bottom section of the surface of the droplet for all $r\leq r_{M}(t)$, and we set  $z_d(r, t) = \infty$ for all other radii. Points on the axisymmetric surface of the drop are parameterised by $\xi(\theta, t) = \xi'(\theta, t)/R_d$ and $\theta$; where $\theta \in [ 0, \pi]$, and $t\geq 0$. Thus the function $z_d$ is given by
\begin{equation}
z_d(r,t) = 
\begin{cases}
\displaystyle \min_{\theta}\bigl\{\cos(\theta)\,\xi(\theta,t)\colon \sin(\theta)\,\xi(\theta,t)=r\bigr\}, 
& r \le r_M(t), \\[1ex]
\infty, 
& r > r_M(t).
\end{cases}
    \label{eqn:sphere_height_definition}
\end{equation}

Given the above-mentioned definition, we can express our KM constraints as
\begin{equation}\label{eqn:KM_press_outside}
    p(r, t) = 0, \qquad r > r_c(t), t \geq 0,
\end{equation}
\begin{equation}\label{eqn:KM_contact_match}
    \eta(r, t) = h(t) + z_d(r, t), \quad r \leq r_c(t), t\geq 0,
\end{equation}
\begin{equation}\label{eqn:no_intersection_condition}
    \eta(r, t) < h(t) + z_d(r, t), \quad  r > r_c(t), t\geq 0,
\end{equation}
and
\begin{equation}
    \label{eqn:km_tangency}
    \partial_r \eta(r, t) = \partial_r z_d(r, t), \quad \text{at} \  r = r_c(t), t \geq 0;
\end{equation}
where the final constraint requires that the drop and the bath share the same tangent cone at the boundary of the contact area. In other words, the KM imposes that the contact angle must be exactly $\pi$. 

\subsection{Model summary}\label{section:model_summary}

Equations (\ref{eqn:km_2}), (\ref{eqn:km_3}), 
(\ref{eqn:vel_cm}), (\ref{eqn:centerofmass_drop}), (\ref{eqn:dirichlet-to-neumann}), (\ref{eqn:drop_governing_adim}) and (\ref{eqn:drop_governing_adim_1}) describe the set of coupled equations that must be solved. 

The axisymmetric impact of a deformable droplet against a fluid bath is thus modelled by the solution to

\begin{subequations}
    \label{eqn:equations_summary}
    \begin{equation}
        \label{eqn:summary_km_2}
        \partial_t \eta =\frac{2}{\Rey} \Delta_H \eta + N\Phi, \quad \forall r\geq0 ;
    \end{equation}
    \begin{equation}
        \label{eqn:summary_km_3}
       \partial_t\Phi = -\Bo \eta + Sr \, Dr \Delta_H \eta + \frac{2}{\Rey} \Delta_H \Phi  - Dr \,p , 
        \forall r\geq0;
    \end{equation}
    \begin{equation}
        \label{eqn:summary_DTN}
        N\Phi(r, t) = \frac{1}{2\pi} \lim_{\epsilon\to 0^+} \int_{\mathbb{R}^2 \setminus B(r, \epsilon)} \frac{\Phi(r, t) - \Phi(s, t)}{|r-s|^3} dA(s);
    \end{equation}
    \begin{equation}
        \label{eqn:h_t}
        \dot{h} = v;
    \end{equation}
    \begin{equation}
        \label{eqn:summary_centerofmass_drop}
        \dot{v} = -\Bo + \frac{3}{4\pi } \int_{r\leq r_c(t)} p(r,t) dA;
    \end{equation}
    \begin{equation}\label{eqn:summary_drop_governing_adim}
        U_l = \dot{A}_l  , \quad \forall l\geq 2;
    \end{equation}
    \begin{equation}\label{eqn:summary_drop_governing_adim_1}
        \dot{U}_l = - l(l+2)(l-1) A_l-2\Ohd (2l+1)(l-1)U_l-lB_l , \quad \forall l\geq 2;
    \end{equation}
\end{subequations}
for all $t\geq0$, where
\begin{equation}
    B_l(t) = \frac{1}{2l + 1} \int_{0}^{\pi} p(\theta, t) P_l(\cos(\theta)) \sin(\theta) d\theta.
    \label{eqn:summary_Legendre_projection}
\end{equation}
These equations are subject to the velocity potential and the free-surface elevation having zero first derivatives at the axis of symmetry, and to the boundary, initial and matching conditions given by
\begin{subequations}
    \label{eqn:restrictions_summary}
    \begin{equation}
        \label{eqn:eta_to_infty}
        \eta \to 0, \quad   \text{as} \ r\to\infty, \forall t\geq 0;
    \end{equation}
    \begin{equation}
        \label{eqn:phi_to_infty}
        \phi \to 0,\quad  |\nabla \phi| \to 0,\quad \text{as} \ \left(r^2+z^2\right)\to\infty, \forall t\geq 0;
    \end{equation}
    \begin{equation}
    \eta(r,0) = 0,\quad\forall r\geq0;
    \end{equation}
    \begin{equation}
        \phi(r,z,0) = 0, \quad\forall r\geq 0, z\leq 0;
    \end{equation}
    \begin{equation}
        h(0) = 1;
    \end{equation}
    \begin{equation}
        v(0) = -\sqrt{\Wed};
    \end{equation}
    \begin{eqnarray}
        r_c(0)=0;
    \end{eqnarray}
    \begin{equation}
        p(r, t) = 0, \quad \forall r > r_c(t), t\geq 0;
    \end{equation}
    \begin{equation}
        \eta(r, t) = h(t) + z_d(r), \quad \forall r \leq r_c(t), t\geq 0;
    \end{equation}
    \begin{equation}
        \eta(r,t)< h(t) + z_d(r),  \quad \forall r > r_c(t), t\geq 0;
    \end{equation}
    \begin{equation}
    \label{eqn:km_tangent_condition}
    \partial_r \eta(r, t) = \partial_r z_d(r, t), \quad \text{at} \  r = r_c(t), \forall t\geq 0;
    \end{equation}
\end{subequations}
where
\begin{equation}
z_d(r, t)= \cos(\theta) \left(1 + \sum\limits_{l = 2}^{\infty} A_l(t) P_l(\cos(\theta)) \right),\quad \forall r \leq r_c(t).
\end{equation}

We note that the system of equations (\ref{eqn:equations_summary}) is a closed system, containing a non-linear subproblem which is to determine the contact radius $r_c(t)$ over which the integral in (\ref{eqn:summary_centerofmass_drop}) is calculated. 

\section{Discretisation and numerical approximation}
\label{sec:discretised_model}
We delineate herein the approximations that convert the above-mentioned continuous system into a discrete set of evolution equations. We shall truncate the series of the surface deformation and velocity series at mode $M+1$; hence, the summations are taken over $l=2,\ldots,M+1$. Spatially, we discretise $r$  uniformly in intervals $\delta_r$, denoting $r_i=i \delta_r$, for $i=0,\ldots K$. At the domain boundary, $r_{max} = K \delta_r$,  we set $\eta(r_{max}, t) = 0$.

We introduce an adaptive time step $\delta^k_t = t^{k+1} - t^{k}$, with $t^1 = 0$, and consider discrete (in time and space if applicable) approximations of all variables of interest, namely $\eta$, $\phi$, $p$, $h$, $v$, $r_c$, $B_l$, $A_l$ and $U_l$. 

We use superscripts to indicate time indices, and subscripts to indicate radial position. 
We thus define $\eta^k_i$ as the approximant to $\eta(i\delta_r,t^k)$, and similarly $\phi^k_i$ and $p^k_i$. 

The vectors $\eta^k$, $\phi^k$, $p^k$ in $\mathbb{R}^{K+1}$ whose components are given by the values for $i=0,\ldots,K$ are represented by the omission of the subscript. We define $h^k$ as the approximant to $h(t^k)$, and similarly for $v^k$ and $r_c^k$. 

Finally, we also define $A^k_l$ as the approximant to $A_l(t^k)$, and similarly for $B_l^k$ and $U^k_l$.

The vectors $A^k$, $B^k$ and $U^k$ in $\mathbb{R}^{M}$ whose components are given by the values for $l = 2,\ldots,M+1$ are represented without subscripts. 

Typically, we set $\delta_r = 1/50$ to guarantee sufficient resolution under the droplet and choose $r_{max}=100$ (noting these are non-dimensionalised with the drop radius), which is large enough to prevent waves from reaching the boundary of the domain over the course of a rebound in order to mimic the behaviour of a bath of infinite lateral extension.

The pressed area $C(t)$ is a circular disk with radius $r_c(t)$ (see figure \ref{fig:schematic_impact}). This variable is resolved to the accuracy provided by the mesh, whose discrete-time approximation is $r_c^k = \delta_r(m(t^k) - 1/2) $ if $m(t^k) > 0$, $r^k_c = 0$ otherwise, where the function $m(t^k)$ takes non-negative integer values and represents the number of contact points of the discretised system. The $1/2$ term added to the expression of $r^k_c$ reflects the assumption that the boundary of the contact area is found exactly at the mid-point between nodes $m(t^k)-1$ and $m(t^k)$.

To find the value of $m(t^{k+1})$ in our implicit finite difference scheme, we formulate a different system of equations for each possible number of contact points $q$; thus generating candidate solutions parametrized by $q$, and set $m(t^{k+1})$ to be equal to the value of $q$ that best satisfies the restrictions imposed by (\ref{eqn:restrictions_summary}). Details of this optimization problem are discussed in the following section.

Time derivatives are approximated using an implicit, variable-step backward difference formula of order 2 (VS-BDF2) scheme \citep[see][]{CelayaEtAl2014,Denner2014}. To this end, we introduce the variable
\begin{equation}
s^k = \delta_t^k/\delta_t^{k-1}    
\end{equation}
and we define
\begin{equation}
    a^k = (1 + 2s^k)/(1+s^k), \ \ 
    b^k = -(1 + s^k), \ \ 
    \text{and} \ \ 
    c^k = \left(s^k\right)^2/ (1 + s^k);
\end{equation}
and approximate all time derivatives by
\begin{equation}
\label{eqn:discretised_Laplacian_def}
    \dot{f}(t^{k+1})\approx\dot{f}^{k+1} \coloneqq \frac{a^k f^{k+1}+b^k f^{k}+c^k f^{k-1}}{\delta_t^k}.
\end{equation}
We use second-order accurate matrix approximations $\bfjfm{N}$ and $\bfjfm{\Delta_H}$ for the spatial operators $N$, given by \eqref{eqn:dirichlet-to-neumann}, and $\Delta_H = \partial_{rr} + (1/r)\,\partial_r$. These linear operators in matrix notation have their derivations in Appendix B of \cite{GaleanoRiosEtAl2017}. For instance, the horizontal Laplacian is approximated by

\begin{equation}
    \left(\bfjfm{\Delta_H} \eta^{k+1}\right)_i \coloneqq 
    \left\{
    \begin{array}{cc}
        \LapH{\eta},
        & i>1;
        \\
        4
        \frac{
        \eta_{i+1}^{k+1}
        -\eta_i^{k+1}
        }{
        \left(\delta_{r}\right)^2},
        & i=1;
    \end{array}\right.
\end{equation}
where we note that $\partial_r \eta = \partial_r \phi =0$ has been imposed at the axis of symmetry.

The far field conditions
(\ref{eqn:eta_to_infty}) are also built into  the approximation when $i = K$, by setting boundary conditions $\eta^{k}_{K+1} =\phi^{k}_{K+1} = 0$, for all $k$.
\newcommand{\BDF}[1]{a^k #1^{k+1}+b^k #1^{k}+c^k #1^{k-1}}

The integral of the contact pressure that appears on (\ref{eqn:summary_centerofmass_drop}) can be represented by a $1\times q$ matrix $\bfjfm{S}(q)$; i.e.,
\begin{equation} \label{eqn:discretised_integral_definition}
    \bfjfm{S}(q) p^{k} \approx \int_{C(t)} p(r,t^k) dA = 2\pi \int_0^{r_c}  p(r, t^k) r dr.
\end{equation}
Similarly, the integral on (\ref{eqn:summary_Legendre_projection}) is approximated via another single-row matrix
\begin{equation}
\label{eqn:discretised_projection}
    B^k_l = \bfjfm{L}(l) p^k \approx \frac{1}{2l+1} \int_0^\pi p(\theta, t^k) P_l(\cos(\theta)) \cdot \sin(\theta) d\theta .
\end{equation}

Matrix $\bfjfm{L}(l)$ transforms a vector of pressure values on the evenly spaced radial mesh of the bath into the coefficients of its (truncated) expansion into axisymmetric spherical harmonics. We highlight that matrix $\bfjfm{L}(l)$ will depend on $A^{k+1}$, because the elevation angle $\theta$ that corresponds to a given point in the radial mesh of the bath changes with $A^{k+1}$. Figure \ref{fig:L_on_Ak} illustrates this dependence. Since $\theta$ depends nonlinearly on the mode amplitudes $A^k$, it is found using a Newton root-finding method. 

\tikzset{every picture/.style={line width=0.75pt}} 
\begin{figure}
    \centering
    \begin{subfigure}{0.48\textwidth}
        \centering
        \begin{tikzpicture}[x=0.75pt,y=0.75pt,yscale=-1,xscale=1]
            
            \draw    (100,124) -- (294.33,123) ;
            \draw    (198.33,116.67) -- (198.33,131.67) ;
            \draw [color={rgb, 255:red, 155; green, 155; blue, 155 }  ,draw opacity=1 ]   (223.33,116.67) -- (223.33,131.67) ;
            \draw [color={rgb, 255:red, 155; green, 155; blue, 155 }  ,draw opacity=1 ]   (248.33,116.67) -- (248.33,131.67) ;
            \draw   (146.17,71.58) .. controls (146.17,42.82) and (169.49,19.5) .. (198.25,19.5) .. controls (227.01,19.5) and (250.33,42.82) .. (250.33,71.58) .. controls (250.33,100.35) and (227.01,123.67) .. (198.25,123.67) .. controls (169.49,123.67) and (146.17,100.35) .. (146.17,71.58) -- cycle ;
            \draw  [dash pattern={on 2.5pt off 2.5pt}]  (198.25,71.58) -- (198.17,123.5) ;
            \draw  [dash pattern={on 2.5pt off 2.5pt}]  (198.25,71.58) -- (223.33,116.67) ;
            \draw    (210.79,93.13) .. controls (207.33,95.97) and (202.33,96.97) .. (198.21,96.54) ;
            
            \draw (182,135) node [anchor=north west][inner sep=0.75pt]  [font=\tiny] [align=left] {$\displaystyle r=0$};
            \draw (212,135) node [anchor=north west][inner sep=0.75pt]  [font=\tiny] [align=left] {$\displaystyle r=\delta _{r}$};
            \draw (197.5,84) node [anchor=north west][inner sep=0.75pt]  [font=\normalsize] [align=left] {$\displaystyle \theta $};
            \draw (95,26) node [anchor=north west][inner sep=0.75pt]   [align=left] {a)};
        \end{tikzpicture}
        \label{fig:undeformed}
    \end{subfigure}
    \hfill
    \begin{subfigure}{0.48\textwidth}
        \centering
        \begin{tikzpicture}[x=0.75pt,y=0.75pt,yscale=-1,xscale=1]
            
            \draw    (100,124) -- (294.33,123) ;
            \draw    (198.33,116.67) -- (198.33,131.67) ;
            \draw [color={rgb, 255:red, 155; green, 155; blue, 155 }  ,draw opacity=1 ]   (223.33,116.67) -- (223.33,131.67) ;
            \draw [color={rgb, 255:red, 155; green, 155; blue, 155 }  ,draw opacity=1 ]   (248.33,116.67) -- (248.33,131.67) ;
            \draw  [dash pattern={on 2.5pt off 2.5pt}]  (198.33,102.83) -- (198.17,123.5) ;
            \draw  [dash pattern={on 2.5pt off 2.5pt}]  (198.33,102.83) -- (223.33,121.97) ;
            \draw    (214.79,115.13) .. controls (210.33,117.97) and (205.33,119.97) .. (198.21,119.54) ;
            \draw   (113.33,102.83) .. controls (113.33,91.33) and (151.39,82) .. (198.33,82) .. controls (245.28,82) and (283.33,91.33) .. (283.33,102.83) .. controls (283.33,114.34) and (245.28,123.67) .. (198.33,123.67) .. controls (151.39,123.67) and (113.33,114.34) .. (113.33,102.83) -- cycle ;
            
            \draw (182,135) node [anchor=north west][inner sep=0.75pt]  [font=\tiny] [align=left] {$\displaystyle r=0$};
            \draw (212,135) node [anchor=north west][inner sep=0.75pt]  [font=\tiny] [align=left] {$\displaystyle r=\delta _{r}$};
            \draw (198,108) node [anchor=north west][inner sep=0.75pt]  [font=\normalsize] [align=left] {$\displaystyle \theta $};
            \draw (95,26) node [anchor=north west][inner sep=0.75pt]   [align=left] {b)};

            \end{tikzpicture}
        \label{fig:deformed}
    \end{subfigure}
    \caption{Dependence of the angle of the point of application of the pressure on the deformation of the droplet. Panels show the angle $\theta$ that corresponds to $r=\delta_r$ when the droplet is (a) undeformed and (b) significantly deformed. The lower height of the centre of mass in panel b) results in a larger value of $\theta$.}
    \label{fig:L_on_Ak}
\end{figure}

Nevertheless, for a given droplet shape, operator $\bfjfm{L}(l)$ is linear in $p^k$. A description of the matrix representation of the linear operators $\bfjfm{S}(q)$ and $\bfjfm{L}(l)$ is given in Appendix \ref{section:Appendix_L}. The equations for the discretised model are given in Appendix \ref{section:discrete_model_summary}.

\section{Results}\label{sec:results}
\begin{table}
\centering
\begin{tabular}{lccc}
\textbf{Parameter}       
& \textbf{Symbol} 
& \textbf{Definition} 
& \textbf{Value}
\\
Impact speed          
& $V_0$          
& --
& $0.5-138\,$cm/s
\\
Droplet radius  
& $R_d$
& --
& $0.02-0.045\,$cm
\\
Density (water)         
& $\rho, \rho_d$        
& --
& $1\,$g/cm$^3$
\\
Surface tension (water) 
& $\sigma, \sigma_d$        
& --                              
& $72.2 \,$dynes/cm
\\
Kinematic viscosity (water)
& $\nu, \nu_d$ 
& --
& $0.978\,$cSt
\\
Density (oil)         
& $\rho, \rho_d$        
& --
& $0.87\,$g/cm$^3$
\\
Surface tension (oil) 
& $\sigma, \sigma_d$        
& --                              
& $18.7 \,$dynes/cm
\\
Kinematic viscosity (oil)
& $\nu, \nu_d$ 
& --
& $2\,$cSt
\\
Gravitational acceleration
& $g$
& --
& $981\,$cm/s$^2$
\\
Bond Number
& $Bo$ 
& $\rho_d g R_d^2/\sigma_d$  
& 
$0.0007-0.0929$
\\
Weber Number
& $\Wed$     
& $\rho_d V_0^2 R_d/\sigma_d$ 
& 
$0.0088-7.76$
\\      
Reynolds Number
& $\Rey$
& $R_d^2/T_d\nu$
& $0.6-310$
\\
Ohnesorge Number
& $\Ohd$ 
& $\nu_d T_d / R_d$  
& 
$0.002-0.0069$
\\
Density ratio
& $\Dr$ 
& $ \rho_d  /\rho$  
& 
$1$
\\
Surface tension ratio
& $ Sr$
& $\sigma_d/\sigma$
& 
$1$
\\
Number of spherical modes
& $M$
&  --
& 
$10-60$
\\
\hline
\end{tabular}
\caption{Parameter ranges explored in this manuscript.\label{tab:parameters}}
\end{table}

\subsection{Validation}

Table~\ref{tab:parameters} lists the full range of physical parameters used across all simulations and experiments. We first verify the predictions of our model by simulating the impact of a water droplet of radius $R_d = 0.035$ cm onto a water bath at velocity $V_0 = 38$ cm/s, corresponding to a characteristic time of $T_d = 0.77$ ms. This scenario matches the experimental case studied by \citep{AlventosaEtAl2023}, enabling a direct comparison between our model’s predictions, and their experimental and numerical results. 

\begin{figure}
    \centering
    \includegraphics[width = \textwidth]{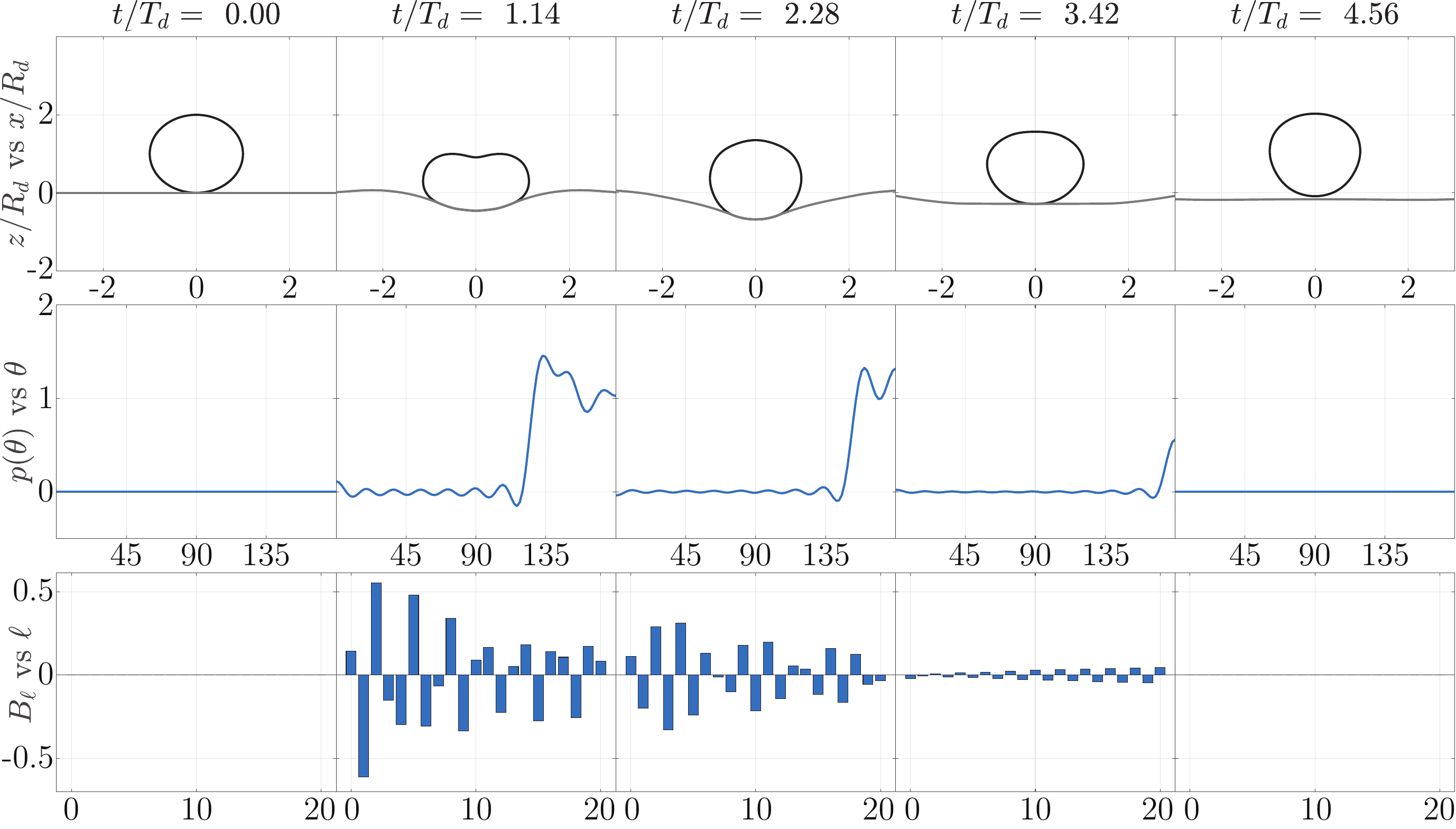}
    \caption{Simulation of a water droplet impacting a water bath at $V_0 = 38$ cm/s, corresponding to $\Wed = 0.7, \Bo = 0.017, \Ohd=0.006$ and $M= 20$. Columns represent snapshots of the impact. From top to bottom, the rows show the spatial reconstruction of the impact, the dimensionless pressure distribution in azimuthal spherical coordinates at the droplet's surface, and the amplitude of the pressure distribution in Legendre's decomposition as a function of the mode number.}
    \label{fig:impact_panels}
\end{figure}

The first row of figure~\ref{fig:impact_panels} shows the droplet shape at equally spaced times during impact. The droplet initially spreads upon contact, reaching a maximum contact area before surface tension drives retraction and powers rebound. The first and last panels correspond to the initial and final moments of contact between the droplet and the bath. The second row of figure~\ref{fig:impact_panels} presents the dimensionless pressure distribution along the droplet surface as a function of azimuthal angle in spherical coordinates. This distribution resembles a step function confined to the contact region, with accuracy determined by the number of modes retained in the simulation. The oscillations observed during spreading arise from truncating the pressure-field expansion. The third row of panels in figure~(\ref{fig:impact_panels}) shows the amplitude attributed to each mode of the Legendre decomposition for the dimensionless pressure distribution. The magnitude of $B_l$ generally decreases with $l$, consistent with the physical interpretation that higher order oscillation frequencies are less important---an interpretation supported by prior results on droplet rebound on non-wetting substrates \citep{GabbardEtAl2025}. However, pressure localization at near detachment excites higher modes, as shown in the column that corresponds to $t/T_d = 3.42$. These higher-frequency excitations are damped in proportion to viscosity and mode number squared (see equation \ref{eqn:summary_drop_governing_adim_1}), ensuring higher modes have little effect on the overall dynamics. Here, $\theta = 0$ corresponds to the droplet’s north pole. If the coordinate system is rotated so that $\theta = 0$ corresponds to the south pole, all coefficients with odd $l$ would change sign, reflecting the odd symmetry of $P_l$. Because higher-order modes become more prominent during spreading, truncation must be chosen carefully to ensure this phase is properly resolved. A sensitivity analysis justifying the truncation level used herein is provided in Appendix \ref{sec:comp_implementation}.

We go on to compare the predicted trajectories of the north pole, south pole, and centre of mass of the droplet against the experimental measurements of \citet{AlventosaEtAl2023} for the same parameters. Note that the north pole of the droplet was experimentally defined as the highest point of the droplet silhouette as captured by a side-view camera, which corresponds to a unique point along the vertical centreline at low $\Wed$, but can correspond to a circumferential ring of points at higher $\Wed$. The droplet was considered to be in contact with the bath when its north pole was below $z=2R_d$; we adopt this criterion to allow for a direct comparison with the prior experimental data, where the exact moment of separation was difficult to determine. The south pole was determined when the droplet was in contact with the surface of the bath and not obscured by the meniscus formed between the bath and its container, as discussed by \citet{AlventosaEtAl2023}. Numerically, these parameters are defined in the same way when post-processing the simulations; the only distinction is that the south pole can be tracked at all times during the simulations. 

\begin{figure}

     \centering
    \includegraphics[width = \textwidth]{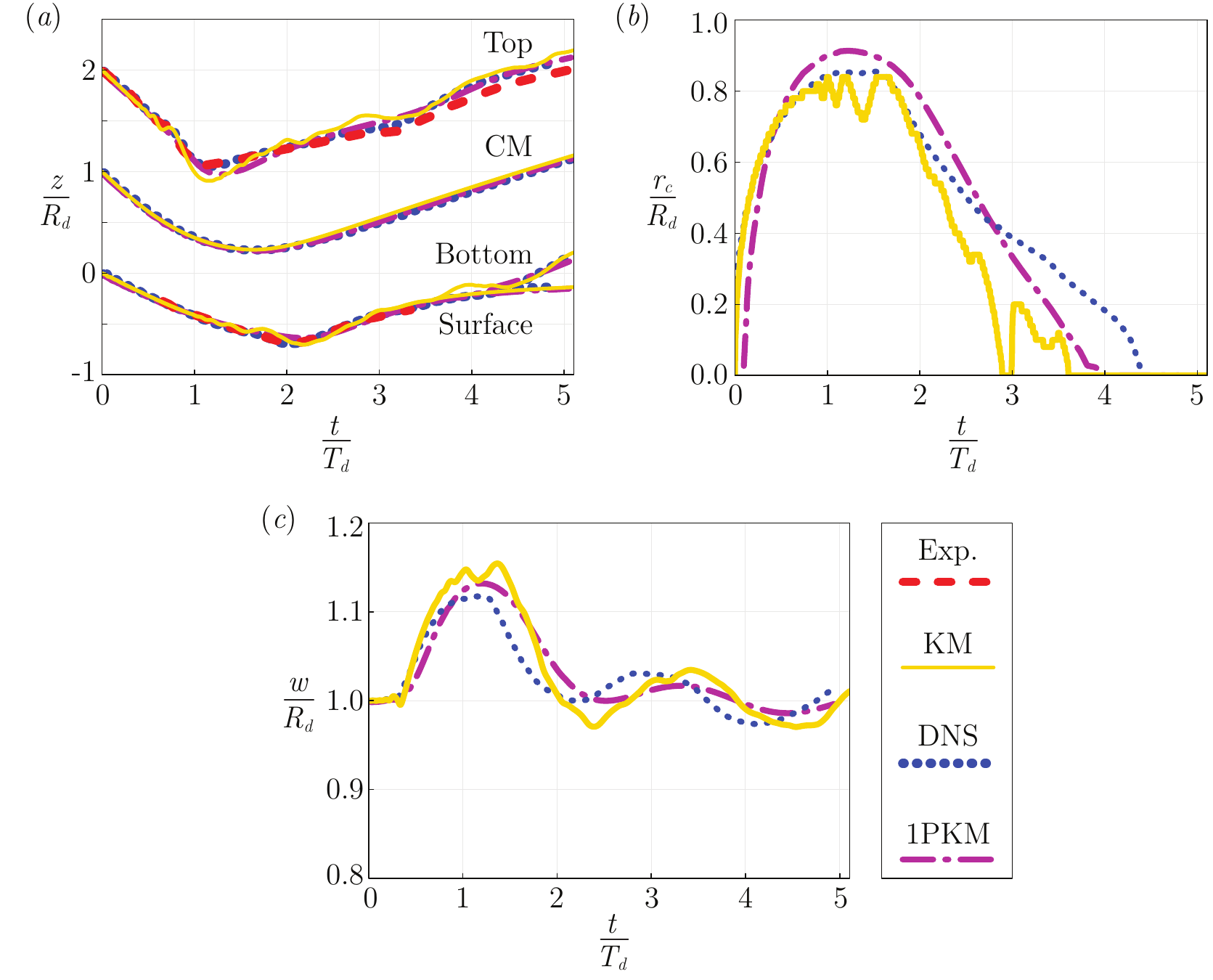}
    \caption{Comparison between model predictions and measurements for a water droplet impacting a water bath at $\Wed = 0.7, \Bo = 0.017, \Ohd=0.006$ and $M = 20$. Results from the full KM model (yellow solid lines) are shown alongside experiments (red dashed lines), DNS simulations (blue dotted lines), and the 1-point KM model predictions (dashed purple lines), both from \cite{AlventosaEtAl2023}. ($a$) Vertical positions of the droplet top, centre of mass, and bottom, together with the bath surface. ($b$) Non-dimensional contact radius. ($c$) Non-dimensional maximum width of the droplet, all plotted as functions of non-dimensional time.}
    \label{fig:drop_trajectory}
\end{figure}

Figure \ref{fig:drop_trajectory}($a$) shows the vertical position of the north pole, south pole, and centre of mass of the droplet, as well as the surface of the fluid bath against nondimensional time. Predictions from the full kinematic match (KM) developed in this manuscript are shown as yellow solid lines, while the DNS simulations, 1-point-kinematic match predictions (1PKM), and experimental measurements from \cite{AlventosaEtAl2023} are shown as a dotted blue line, dash-dotted purple line, and red dashed line, respectively. The full KM model captures the dynamics with close agreement to both prior models and experimental data. In particular, it reproduces the motion of the droplet’s center of mass with excellent accuracy, an essential feature for reliably predicting rebound characteristics such as the coefficient of restitution. 

\begin{figure}
   
        \centering
    \includegraphics[width = \textwidth]{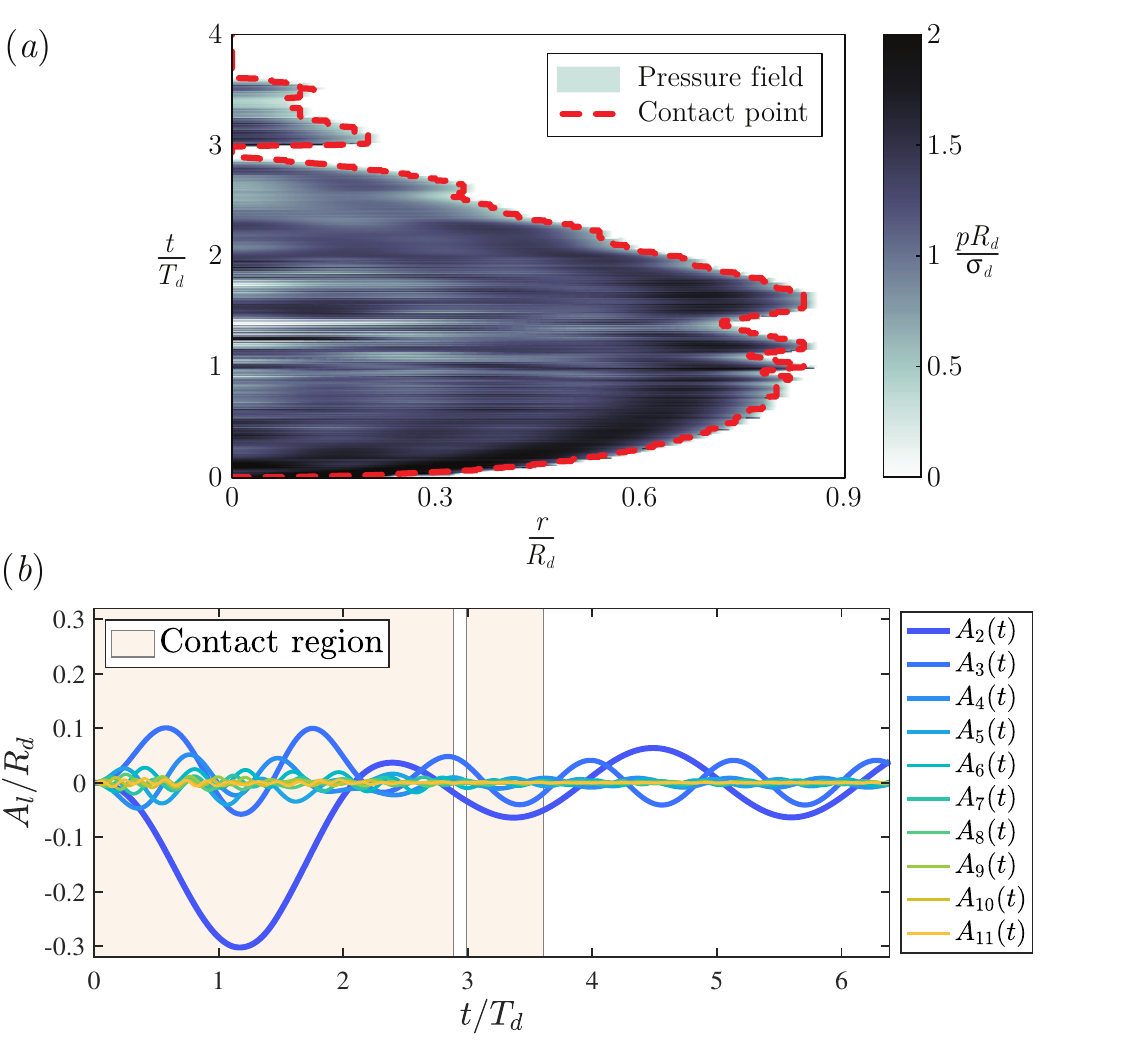}
    
    \caption{($a$) Evolution of the dimensionless pressure field along the pressed area in cylindrical coordinates. The red dashed line represents the extent of the numerical contact radius. ($b$) Evolution of the dimensionless mode amplitudes $A_l$ for the first modes of oscillation. Line thicknesses are inversely proportional to the mode number. The shaded area shows the timespan when the droplet and bath are in contact. Both panels correspond to the same impact as in figure \ref{fig:impact_panels}.}
    \label{fig:pressure_field_modes_amplitudes}
\end{figure}

Figure \ref{fig:drop_trajectory}($b$) compares the predicted evolution of the instantaneous contact radius of the droplet amont the full KM model, DNS simulations, and 1PKM model. Overall, the methods agree reasonably well with the full KM model attaining a similar maximum contact radius as DNS, and a similar time between first and last contact when compared with the 1PKM model. Two distinct features arise in the full KM predictions. First, as the droplet approaches its maximum spread, small oscillations in the radius of the contact area are visible. These oscillations result from surface waves propagating up and down the droplet, as a consequence of the moving pressure field, and can be clearly seen in the rebound animation in the supplementary material. Oscillations of the location of the contact boundary and of the pressure field (see Figure \ref{fig:pressure_field_modes_amplitudes}(a)) are not reported in experimental observations or DNS results. In experiments, the pressure field cannot be measured directly, and the definition of contact usually requires a cut-off criterion due to the intervening air layer. This criterion determines what is resolved and may prevent oscillations of the type predicted by our model from appearing in the data. DNS results were obtained by volume-of-fluid methods where interfaces are reconstructed from approximate level sets, which may also lead to damping of oscillations. Given the small slopes of the surface of the droplet at these locations, even small differences in the cut-off thickness of the air-layer can absorb the oscillations that our model predicts. We also note that some oscillations in the pressure are seen in the sharp interface lubrication layer model for solid rebounds of \citet{phillips2024lubrication}. More interestingly, these oscillations are not observed in the 1PKM. This discrepancy is expected, as the 1PKM model excites a more restrictive set of surface modes (namely those used in the prescribed form of the pressure field). The other noticeable difference in the prediction of the full KM is the brief lift-off period just before rebound, around $t/T_d = 3$. This effect also results from neglecting the intervening air layer, and assuming direct liquid–liquid contact. While this simplification is consistent with previous studies \citep{AlventosaEtAl2023}, and supported by the results presented here, it produces a short interval where the droplet surface rises above the bath by a height that is less than the typical air-film thickness ($\sim 100$ nm; \citep{CouderEtAl2005Noncoalescence}).

\begin{figure}
    \centering
    \includegraphics[width=0.85\linewidth]{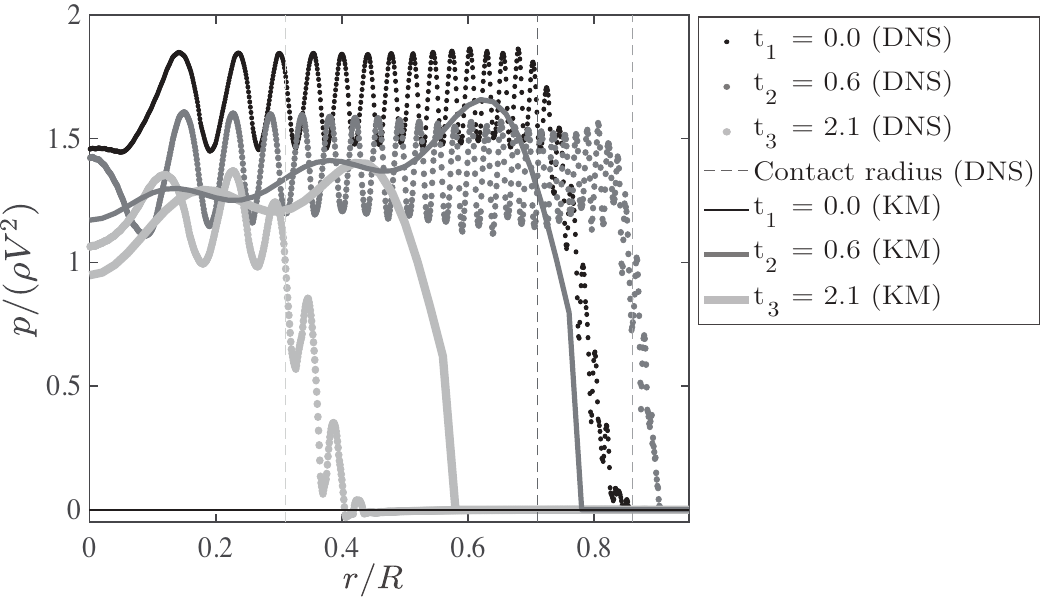}
    \caption{Normalised pressure distribution versus dimensionless cylindrical coordinate $r/R_d$ along the surface of the bath. Solid lines denote the KM model, and points represent DNS results from \cite{AlventosaEtAl2023}. Plots correspond to three (dimensionless) times: $t_1 = 0.0$, $t_2 = 0.6$, and $t_3 = 2.1$, which are thickness and colour coded. Vertical dashed lines indicate the contact radii extracted from the DNS. Simulation parameters correspond to $\Wed = 0.7$, $Bo = 0.017$, and $\Ohd = 0.006$ as defined in Figure \ref{fig:impact_panels}.}
    \label{fig:DNS_pressure_comparison}
\end{figure}

Figure~\ref{fig:drop_trajectory}($c$) compares the non-dimensional maximum width of the droplet predicted by the full KM model, DNS simulations, and the 1PKM model. Unlike in panel ($b$), where the contact radius is restricted to discrete values by the mesh, the width in the full KM model is obtained from the spectral representation of the droplet. The Newton method is used to locate the vertical tangent plane at the droplet surface, allowing the width to vary continuously. All three approaches show good agreement, particularly in the maximum width reached during the initial spreading phase. The oscillation periods predicted by the full and 1PKM models are mutually consistent, although both are slightly longer than those from DNS. Since these differences primarily influence the post-rebound droplet shape, they have little effect on accurately capturing rebound metrics, as will be shown in the next section.

A key strength of the full KM model developed here is its ability to directly predict the pressure distribution across the contact area, whereas in the 1PKM model the spatial form of the distribution must be prescribed \citep{AlventosaEtAl2023}. Figure~\ref{fig:pressure_field_modes_amplitudes}($a$) presents the dimensionless pressure field for the water drop impact studied in figures~\ref{fig:impact_panels} and \ref{fig:drop_trajectory}, with the $x$- and $y$-axes denoting time and radial distance from the axis of symmetry, respectively. The highest pressures occur near the boundary of the contact region. Importantly, the spatio-temporal evolution of the contact pressure is highly non-trivial, underscoring the advantage of the full KM model in capturing these dynamics. Figure \ref{fig:pressure_field_modes_amplitudes}($b$) shows the evolution of the first ten mode amplitudes. Higher order modes are not plotted as their amplitudes were negligible compared with those presented in the figure. The first mode of oscillation $A_2$ is dominant, having an absolute peak at about $t/T_d \approx 1$ that roughly corresponds to the time of maximum contact area. The average absolute amplitude of a mode $A_\ell$ is approximately in inverse proportion to its index $\ell$. 

We compare the radial pressure profiles computed by the KM model against DNS results from \cite{AlventosaEtAl2023} in figure \ref{fig:DNS_pressure_comparison}. The profiles are shown for the three dimensionless times analysed by \citet{AlventosaEtAl2023}: start of impact ($t_1=0$), maximum spreading ($t_2=0.6$), and retraction ($t_3=2.1$). The KM model demonstrates good quantitative agreement with the fully resolved DNS, capturing both the pressure magnitude and the sharp drops in pressure near the moving contact line. We highlight that $t_1 = 0$ was defined as the time at which contact would be starting (single point of contact) in \citet{AlventosaEtAl2023} for an airless model. For this time, the KM model predicts zero pressure, but the DNS already shows an effect of the droplet presence. The overall comparison supports the case for the use of the KM method, especially when considering the simplifying assumptions introduced here, as well as the gains in computational cost with respect to DNS.

\subsection{Rebound metrics} \label{subsec:ReboundMetrics}

To compare our simulation results with experimental values, we measure the contact time $t_c$ as the difference between the times when the north pole of the droplet crosses $z=2R_d$ during descent and rebound; that is, following the experimental convention. Moreover, the coefficient of restitution $\alpha$ is defined as the square root of the ratio of the rebound energy $E_r$
 to impact energy $E_0$:
\begin{equation}
\alpha = \sqrt{\frac{E_{r}}{E_{0}}} = \sqrt{\frac{g(h_{r}-R_d) +  0.5V_{r}^2}{0.5 V_{0}^2}},
\end{equation}

where $h_r = h(t = t_c)$ represents the center of mass height at rebound, while $V_0$ and $V_r$ represent the center of mass velocities at impact and rebound, respectively. This definition sets the potential energy of the droplet at $t=0$ to zero and allows for a direct comparison with experiments. We define the maximum penetration depth $\delta$ as minus the lowest height of the south pole during impact, as measured from $z=0$, consistent with the experimental definition used by \cite{AlventosaEtAl2023}.

\begin{figure}
        \centering
    \includegraphics[width = \textwidth]{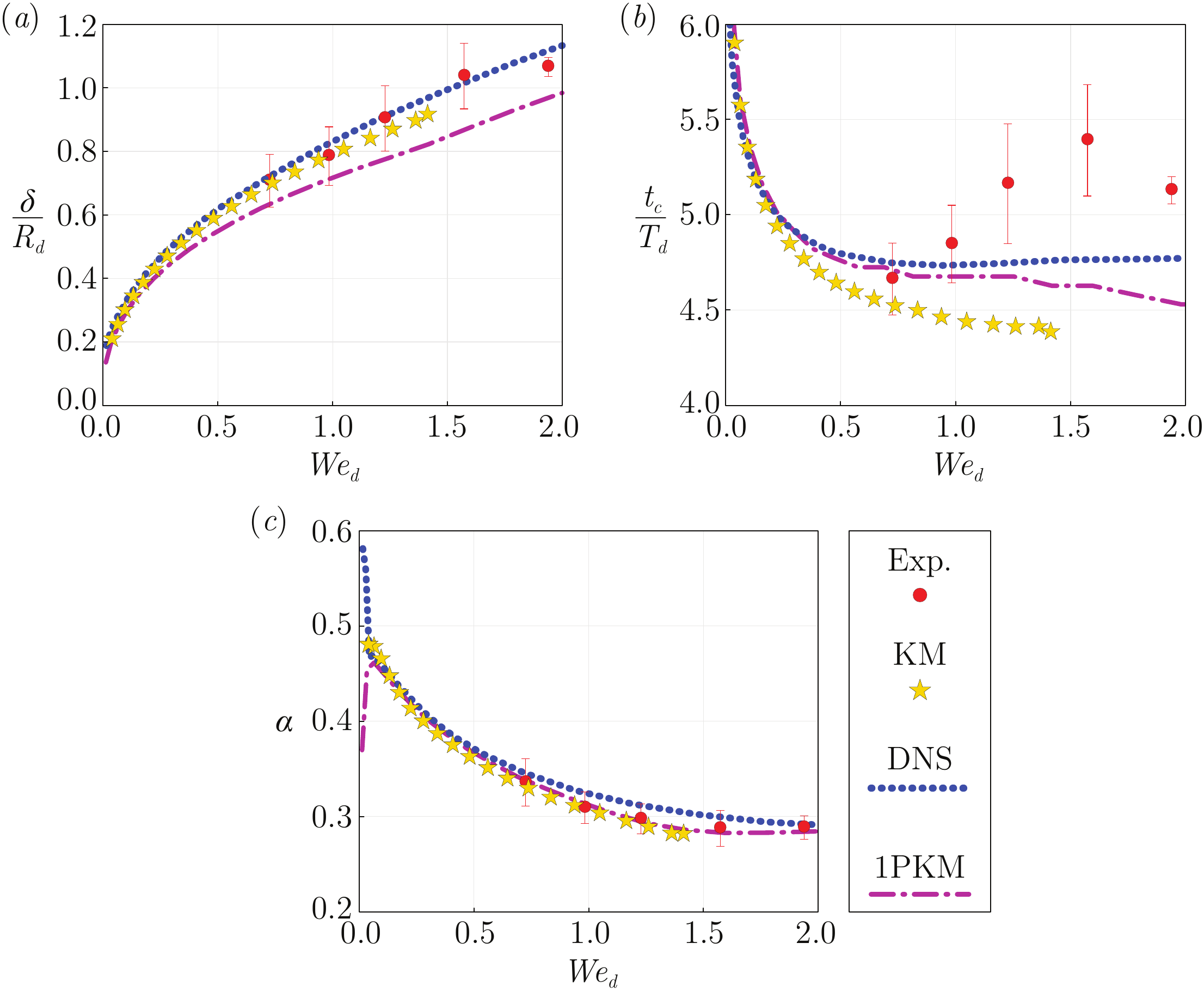}
    \caption{Non-dimensional parameters as a function of  $\Wed$. ($a$) Maximum penetration depth, ($b$) contact time and ($c$) coefficient of restitution for a droplet with the same non-dimensional parameters as in figure \ref{fig:impact_panels}. Experimental results are red circles, predictions from the full kinematic match model are yellow stars, DNS results are blue dotted lines and predictions from the 1-point kinematic match from \citep{AlventosaEtAl2023} are purple dash-dotted lines.}
    \label{fig:contact_time_maxdef_alfa}
\end{figure}


 \begin{figure}
    \centering
    \includegraphics[width = \textwidth]{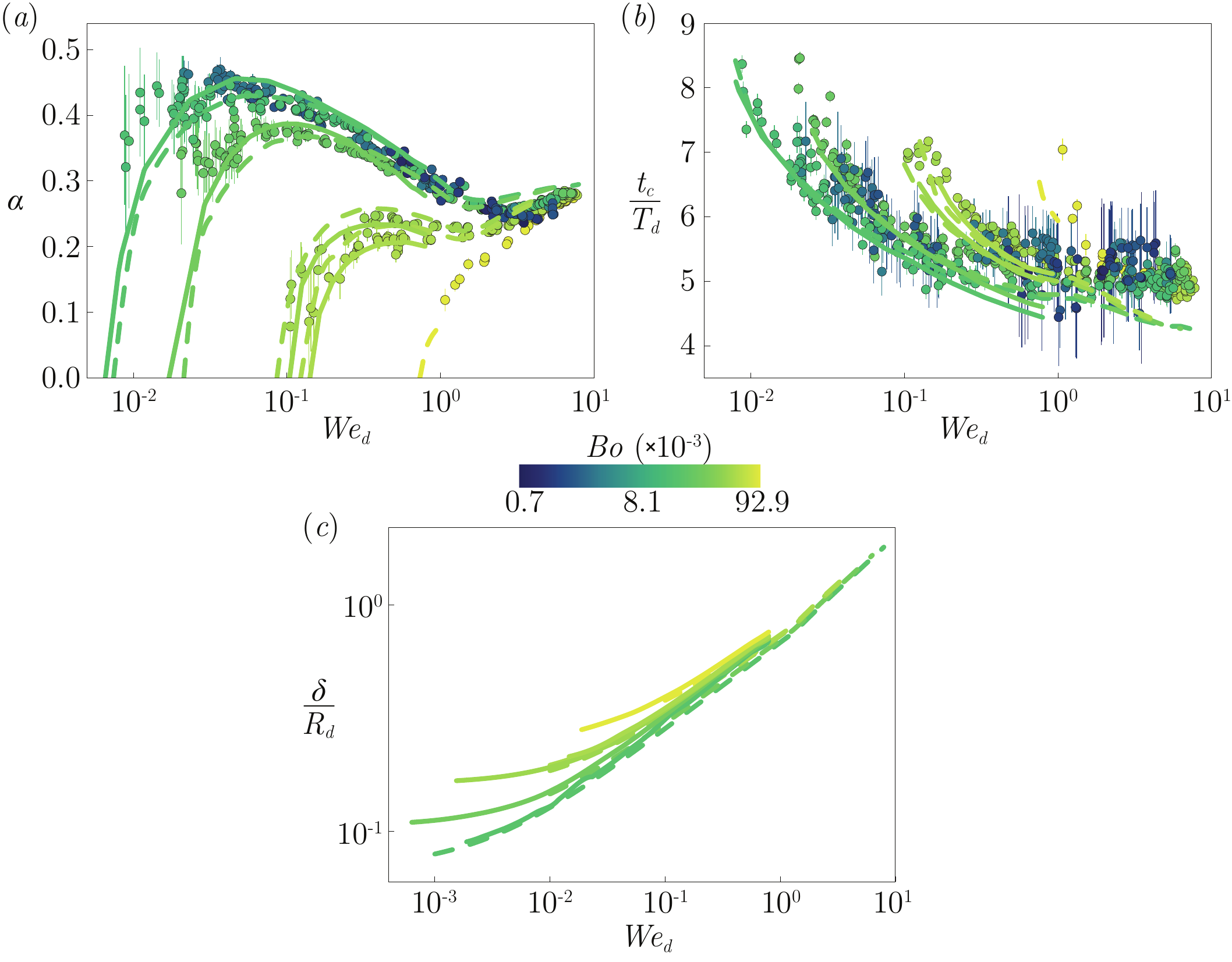}
    \caption{($a$) Coefficient of restitution $\alpha$, ($b$) dimensionless contact time $t_{c} / T_{d}$, and ($c$) dimensionless maximum deflection $\delta / R_{d}$ versus Weber number $\Wed$ for drops impacting a bath of the same liquid. Circles are experimental results, and dashed and solid lines are predictions from the 1-point and full kinematic match models, respectively. The color bar maps $\Bo$ to color on a logarithmic scale, and $\Ohd$ ranged from 0.0234 to 0.0351 in the simulations.}
    \label{fig:FullComparisions}
\end{figure}

We further compare predictions from the full KM model and 1-point KM model derived by \citet{AlventosaEtAl2023} with new experiments. In our experiments, we generated submillimetre-sized droplets  and tracked their motion as they impact a bath of the same liquid. The drop trajectory and shape were determined from high-speed recordings, and we measured the contact time $t_c$ and coefficient of restitution $\alpha$ for $\Wed$ down to $10^{-2}$. Below this threshold, droplets either coalesced with the bath or transiently floated, depending on $\Bo$. Here, we define either outcome as a non-bounce event, where a coalescing drop merges with the bath during impact or its initial rebound, while a transiently floating drop persists longer on the surface: it never rises during the first rebound and only coalesces during a later oscillation on the bath. 

Figure~\ref{fig:FullComparisions} shows the coefficient of restitution $\alpha$, dimensionless contact time $t_{c} / T_{d}$, and dimensionless maximum penetration depth $\delta / R_{d}$ versus $\Wed$ for a range of $\Bo$ values, indicated by color on a logarithmic scale. Experiments are denoted by circles while predictions from the 1-point KM and full KM model are shown as dashed and solid lines, respectively. As most experiments involve low-$\Wed$ impacts, the maximum vertical deformation is obscured in experiments and only model predictions are compared. Error bars are associated with experimental uncertainties, as discussed in Appendix~\ref{app:experiment}.

In Figure~\ref{fig:FullComparisions}($a$), three trends in $\alpha$ are observed. When $\Wed \gtrsim 1$, the coefficient of restitution $\alpha$ increases with $\Wed$, approaching a $\Bo$-independent value of $\approx0.3$, consistent with prior literature \citep{bach2004coalescence,wu2020small,AlventosaEtAl2023}. Below $\Wed \approx 1$, $\alpha$ increases as $\Wed$ decreases over a range that widens as $\Bo$ decreases. This increasing trend ceases once $\Bo$ effects become significant, causing $\alpha$ to roll off to zero, i.e., drops transition from bouncing to transiently floating on the interface. This rolloff mirrors that recently reported for droplets bouncing on a rigid, non-wetting substrate \citep{GabbardEtAl2025}, but here both the onset of rolloff and the cessation of rebound occur at higher $\Wed$ due to the deformable substrate. For the smallest $\Bo$ experiments (dark blue), there is no roll off as bouncing drops coalesce with the bath at low $\Wed$ rather than transiently float. As the kinematic match framework does not consider the air layer, and thus can not capture coalescence, we only simulate the $\Bo$ associated with a bounce-to-float transition in experiment.  The 1PKM and full KM models capture all three trends in $\alpha$, with exceptional accuracy in predicting the transition from bouncing to floating. Only predictions for the 1PKM are provided at the largest $Bo$ (yellow) as the high deformation associated with the significantly larger droplet size in this regime prevented full KM simulations from converging.

Figure~\ref{fig:FullComparisions}($b$) shows the dimensionless contact time $t_{c} / T_{d}$ versus $\Wed$. For $\Wed \gtrsim 1$, the contact time is $\Wed$-independent, but increases as $\Wed$ decreases in the low-$\Wed$ limit, similar to drop rebound on a rigid, non-wetting substrate \citep{GabbardEtAl2025}. The contact time begins rising at higher $\Wed$ as $\Bo$ increases. The 1PKM and full KM models accurately predict these trends and show quantitative agreement except for the largest $\Wed$.  Figure~\ref{fig:FullComparisions}($c$) compares predictions for the maximum deflection $\delta / R_{d}$ for the 1-point and full KM models. The model predictions are nearly identical across all $\Wed$. The initial decrease in deflection transitions to nearly constant deflection in the low-$\Wed$ limit, where the asymptotic value depends solely on $\Bo$. 

\section{Discussion}\label{sec:Discussion}
This is the first work to implement the kinematic match method to model a deformable impactor colliding with a liquid bath. Our results show that this method accurately reproduces our experimental results as well as experimental and numerical results found in the literature with significantly reduced computational costs compared with the direct numerical simulations of \citet{AlventosaEtAl2023}. A typical run of the full kinematic match takes 4 CPU hours, representing a twelvefold reduction compared with the average DNS run time reported by \cite{AlventosaEtAl2023}. 

Our results demonstrate the usefulness and versatility of the kinematic match approach when coupled with linearised fluid models, particularly for problems involving complex contact dynamics in the low-Weber-number limit. This capability is especially relevant for improving models of superwalkers \citep{ValaniEtAl2019} and other bouncing droplet phenomena where droplet deformation is essential. Furthermore, it naturally applies to problems where the liquid bath is replaced with an elastic medium \citep{AgueroEtAl2022}. These findings, combined with the fact that the contact conditions imposed by \eqref{eqn:KM_press_outside}-\eqref{eqn:km_tangency} are largely model-agnostic, encourage further exploration of the KM method to many more applications, including its use with nonlinear fluid models, extension to non-axisymmetric cases, and use within variational formulations (e.g., for problems that involve more complex geometry). Indeed, provided there is no actual wetting (in the case of a droplet on a solid surface or a solid impactor on a liquid surface), coalescence, or droplet breakup, the approach presented here is, in principle, generalizable to almost any impact regime.

A significant insight from the present work is that the non-linear problem of determining the pressed area between two deformable impactors can be successfully addressed by dividing it into two iterative cycles. The external cycle explores the possible shapes of one impactor, while the internal cycle determines the extent of the pressed area. A remaining avenue for exploration is to reduce the iterative nature of the solution by parameterising the geometry and solving a nonlinear system of equations in which the geometric nonlinearities are absorbed into algebraic nonlinearities of the parameterisation.

Lastly, this study offers new insights into the physics of droplet rebound on a liquid bath, enabling direct comparisons between impacts on rigid and liquid substrates at low Weber number. For small droplets on a liquid bath, the energy recovered during rebound remains nearly constant at moderate $\Wed$ but increases when $\Wed \lesssim 1$ for low $\Ohd$ \citep{zhao2011transition,wu2020small,AlventosaEtAl2023}. We further show that at even lower $\Wed$, $\alpha$ exhibits a roll-off that depends critically on $\Bo$, analogous to behaviour observed for droplets impacting non-wetting substrates \citep{GabbardEtAl2025}. However, our experiments and model reveal that a deformable substrate shifts this rolloff to significantly higher $\Wed$, suggesting that substrate compliance may limit droplet transport via bouncing. The contact time $t_{c}$ scales with $T_{d}$ \citep{zhao2011transition,AlventosaEtAl2023} similar to droplets bouncing on a rigid substrate \citep{richard2002contact}, but shifted towards longer interaction times. At lower $\Wed$, the contact time increases, with $t_{c} \rightarrow \infty$ for large $\Bo$ droplets that can transiently float on the bath \citep{CouderEtAl2005Noncoalescence}, while for low $\Bo$, contact time progressively increases until it exceeds the drainage time for the intervening air film, leading to coalescence. These results provide the most detailed characterization to date of submillimetre-sized droplet impacts in the low-$\Wed$ regime. The exceptional accuracy of the kinematic match model in this limit makes it a powerful predictive tool for a range of timely applications, including the spread of cough ejecta \citep{bourouiba2021fluid} and agricultural sprays \citep{makhnenko2021review} along liquid surfaces. \\

{\small \noindent \textbf{Supplementary movie}. Supplementary movie available at \href{https://doi.org/10.1017/jfm.2026.11291}.}

{\small
 \noindent\textbf{Funding.} C.A.G-R. gratefully acknowledges the support of EPSRC project EP/P031684/1. C.A.G-R. and P.A.M gratefully acknowledge the support of EPSRC project EP/N018176/1. C.R. acknowledges FAPESP grants Sprint-17/50295-8 and 2023/07076-4. C.T.G. gratefully acknowledges the Hope Street Fellowship for financial support. D.M.H. gratefully acknowledges the financial support of the National Science Foundation (NSF CBET-2123371).\\

 \noindent\textbf{Data availability.} The data that support the findings of this study are openly available in GitHub at \href{https://github.com/harrislab-brown/km-droplet-onto-bath}{https://github.com/harrislab-brown/km-droplet-onto-bath}. \\

 \noindent\textbf{Declaration of Interests.}
The authors report no conflict of interest. \\ }

\appendix

\section{Operators \(\bfjfm{L}(\ell)\) and $\bfjfm{S}(q)$}
\label{section:Appendix_L}
\subsection{Construction of $\mathbf{L}(\ell)$}
In the main text, we introduced the matrix operator \(\bfjfm{L}(\ell)\) that projects the discrete pressure values at a given time
\(p(\theta_i) :=p(\theta_i, t)\) in the droplet onto the Legendre‐mode coefficients \(B_\ell\).  In this appendix, we derive \(\bfjfm{L}(\ell)\), rendering $B_\ell$ as explicit linear combinations of the nodal pressures \(p(\theta_i)\).  We begin by recalling the continuous projection
\begin{equation}
B_\ell \;=\;\frac{1}{2\ell+1}\int_{0}^{\pi} p(\theta)\,P_\ell\bigl(\cos\theta\bigr)\,\sin\theta\,d\theta.
\end{equation}

We split the integral to be approximated and use the change of variables formula to get:
\begin{equation}
    B_l = \frac{1}{2l+1} \int_0^\pi p(\theta) P_l(\cos(\theta)) \cdot \sin(\theta) d\theta
    =
    \frac{1}{2l+1} \sum_{i=0}^{q} \int_{\cos(\theta_i)}^{\cos(\theta_{i+1})} \bar{p}(x) P_l(x) dx,    \label{eqn:numerical_implementation_projection}
\end{equation}
where $\bar{p}(x) = p(\arccos(x))$. Under the assumption that $p$ is a piecewise linear function of $x = \cos(\theta)$ on the original spherical coordinate system, we employ a piecewise linear representation, 
\begin{equation}
  \bar{p}(x) = ax + b = p(\theta_{i+1})\frac{x - \cos(\theta_i)}{\cos(\theta_{i+1}) - \cos(\theta_i)} + p(\theta_{i})\frac{x - \cos(\theta_{i+1})}{\cos(\theta_{i}) - \cos(\theta_{i+1})} \ ,  
\end{equation}
where $a$ and $b$ are chosen so that $p(\cos(\theta_i))$ is equal to the prescribed query values at $\cos(g(i \delta_r, z^{k, j})) := \cos(\theta_i)$, where $g$ is a function that translates from cylindrical coordinates to spherical coordinates. We can integrate each term in \ref{eqn:numerical_implementation_projection} exactly, resulting in  a linear combination of the pressure values at the endpoints of each subinterval:
\begin{equation}
    \int_{\cos(\theta_i)}^{\cos(\theta_{i+1})} \bar{p}(x) P_l(x) dx
    =
    a \int_{\cos(\theta_i)}^{\cos(\theta_{i+1})} x P_l(x) dx 
    +
    b \int_{\cos(\theta_i)}^{\cos(\theta_{i+1})} P_l(x) dx.
\end{equation}

We compute these integrals exactly since the antiderivatives $F_l(x) = \int P_l(x) dx$ and $G_l(x) = \int x P_l(x) dx$ satisfy the following Bonnet-type recurrence relations:

\begin{subequations}
    \begin{equation}
        (2l+1) P_l(x) = \frac{d}{dx} \bigl( P_{l+1}(x) - P_{l-1}(x)) \implies F_l = \frac{P_{l+1} - P_{l-1}}{2l+1}, 
        \label{eqn:bounnet_1}
    \end{equation}
    \begin{equation}
        (l+1) P_{l+1}(x) = (2l+1) x P_l(x) - l P_{l-1}(x) \implies G_l = \frac{(l+1)F_l + l F_{l-1}}{2l+1} ,
        \label{eqn:bounnet_2}
    \end{equation}
\end{subequations}

The estimation of $B_l$ then becomes

\begin{equation}
    B_l \approx \frac{1}{2l+1} \sum_{i=0}^{q} \left(a_i G_l\bigg|_{\cos(\theta_i)}^{\cos(\theta_{i+1})} + b_i  F_l\bigg|_{\cos(\theta_i)}^{\cos(\theta_{i+1})} \right) \ ,
\end{equation}
where
\begin{equation}
    a_i = \frac{p(\theta_{i+1}) }{\cos(\theta_{i+1}) - \cos(\theta_i)} + \frac{p(\theta_{i})}{\cos(\theta_{i}) - \cos(\theta_{i+1})} \ ,
\end{equation}
\begin{equation}
    b_i = p(\theta_{i+1})\frac{ - \cos(\theta_i)}{\cos(\theta_{i+1}) - \cos(\theta_i)} + p(\theta_{i})\frac{ - \cos(\theta_{i+1})}{\cos(\theta_{i}) - \cos(\theta_{i+1})} \ .
\end{equation}

Approximating the Legendre modes from the pressure values thus requires computing the value of $P_l(\cos(\theta_i))$, for all $ l = 2, \dots M+1$, for all $ i = 0, \dots q$. This is done in linear time as a function of $q$ and $M$, the only assumption being the linearity of the pressure as a function of $x = \cos(\theta)$. It can be noted that recurrence relations exist for antiderivatives of the form $\int x^n P_l(x) dx$ too, thus allowing for integration of the pressure assuming piecewise polynomial behaviour of arbitrary order.

\subsection{Construction of \(\mathbf{S}(q)\)}

We recall that $\bfjfm{S}(q)$ is defined  to approximate the integral of the pressure in cylindrical (bath) coordinates,

\begin{equation}
    \bfjfm{S}(q) p^{k} \approx 2\pi \int_0^{r_c}  p(r, t^k) r dr,
\end{equation}
where pressure on the contact area between the drop and the bath has been reduced to a one-dimensional function $p(\cdot, t^k)$ because of our axisymmetric assumptions, and $t^k$ is a constant. The area of integration was originally $C(t^k)$, a two-dimensional (2-D) flat disk, turned into a one-dimensional (1-D) integral. 

We have query points $p^k_j := p(j \delta_r, t^k)$, for $j = 0, \dots, q-1$, where $r_c = (q - 1)\delta_r + \frac{\delta_r}{2}$. We split the integral into subintervals,
\begin{equation}
    \bfjfm{S}(q) p^{k} \approx 2\pi \left( \int_{(q-1)\delta_r }^{(q-1/2)\delta_r}  p(r, t^k) r dr + \sum_{i=0}^{q-2}\int_{i \delta_r}^{(i+1)\delta_r}  p(r, t^k) r dr\right).
\end{equation}

Under the assumption that the pressure distribution is a linear interpolation of the query points from $0$ to $r_c$, we calculate the integral in each subinterval as a linear combination of the query points. For example,

\begin{equation}
    \int_0^{\delta_r} \left(p^k_0 + \frac{p^k_1 - p^k_0}{\delta_r} r\right) r dr
    = p^k_0 \frac{\delta_r^2}{2} + \left(\frac{p^k_1 - p^k_0}{\delta_r}\right) \frac{\delta_r^3}{3}
    =
    \delta_r^2 \left(\frac{p^k_0}{6} + \frac{5p^k_1}{6}\right).
\end{equation}

By adding up the contribution of the integral in each subinterval, we obtain the following vector:
\begin{equation}
    \bfjfm{S}(q) = \pi \delta_r^2\left[ \frac{1}{3}, 2, 4, 6, \dots, 2(q-2), \frac{3}{2}(q-1) - \frac{1}{4}\right].
\end{equation}

\section{Discretised model}\label{section:discrete_model_summary}

We formulate the system for a fixed unknown value $q=m(t^{k+1})$ in what follows. Hence, the system of equations (\ref{eqn:equations_summary}) takes the following discretised form for all $k \geq 1$:
\begin{subequations}
    \label{eqn:discretised_equations_summary}
    \begin{equation}
        \label{eqn:summary_discretised_km_2}
        \BDF{\eta} = \delta_t^k \left(
        \frac{2}{\Rey} \bfjfm{\Delta_H}\eta^{k+1}
        + \bfjfm{N}\phi^{k+1}\right);
    \end{equation}
    \begin{align}
        \label{eqn:summary_discretised_km_3}
       &\BDF{\phi} = \notag \\ 
       &\delta^k_t \left(-\Bo \eta^{k+1} + Sr \, Dr \bfjfm{\Delta_H}\eta^{k+1} +
       \frac{2}{\Rey} \bfjfm{\Delta_H}\phi^{k+1}  - Dr \,p^{k+1}\right) ;
    \end{align}
    
    \begin{equation}
        \label{eqn:summary_discretised_centerofmass_drop}
        \BDF{v} = \delta^k_t \left(-\Bo + \frac{3}{4\pi } \bfjfm{S}(q) p^{k+1}\right);
    \end{equation}
    \begin{equation}
        \label{eqn:discrete_h_t}
        \BDF{h} = \delta^k_t v^{k+1};
    \end{equation}
    \begin{equation}\label{eqn:summary_discretised_drop_governing_adim}
        \BDF{A_l} =  \delta^k_t U^{k+1}_l  , \quad \forall l = 2, 3, \dots M+1;
    \end{equation}
    \begin{align}
    \label{eqn:summary_discretised_drop_governing_adim_1}
        \BDF{U_l} &= \\ \nonumber
        \delta^k_t \left(- l(l+2)(l-1) A^{k+1}_l - 2\Ohd (2l+1)(l-1) U^{k+1}_{l}-lB^{k+1}_l \right)&, \forall l = 2, 3, \dots M+1.
    \end{align}
    \begin{equation}
    \label{eqn:discrete_Legendre}
    B^k_l = \bfjfm{L}(l)  p^k, \quad \forall l = 2, \dots, M+1.
\end{equation}
\begin{equation}
    \label{eqn:zero_pressure_discretised}
        p^k_i = 0, \quad \forall i > q.
    \end{equation}
\end{subequations}

Subject to discretised initial and boundary conditions, and the discretised approximations of the system of restrictions (\ref{eqn:restrictions_summary}). In particular, for all $k\geq 1$, we have

\begin{subequations}
    \label{eqn:summary_discretised_restrictions}
    \begin{equation}
        \label{eqn:discrete_contact_condition}
        \eta^k_i = h^k + z_d(i  \delta_r, t^k), \quad \forall i \leq q;
    \end{equation}
    \begin{equation}
        \label{eqn:discrete_condition_no_overlay_drop}
        \eta^k_i< h^k + z_d(i \delta_r, t^k),  \quad \forall i > q;
    \end{equation}
    \begin{equation}
    \label{eqn:summary_km_tangent_condition}
    \frac{\eta^k_{q+1} - \eta^k_{q}}{\delta_r} = \partial_r z_d((q - 1/2) \delta_r, t^k),\quad  \text{if } q > 0.
    \end{equation}
    
\end{subequations}

Before making further substitutions, we recast the system of equations (\ref{eqn:summary_discretised_km_2})-(\ref{eqn:summary_discretised_drop_governing_adim_1}) into two coupled matrix forms

\begin{equation}
\label{eqn:matrix_form_uncoupled}
    \bfjfm{Q}(\delta_t^k) W^{k+1} = F, \quad \bfjfm{R}(\delta^k_t) Y^{k+1} = G;
\end{equation}
where
\renewcommand{\arraystretch}{1.8}

\setlength{\arraycolsep}{4pt}
 \begin{equation}
     \label{eqn:bath_matrix_form_1}
     \bfjfm{Q}(\delta_t^k)
     =
     \left[
    \begin{array}{ccccc}
        a^k\left(\bfjfm{I}- \dfrac{ 2}{\Rey} \bfjfm{\Delta_H}\right) 
        & -\delta_t^{k}\bfjfm{N}
        & \mathbf{0}_{K\times K}
        & \mathbf{0}_{K\times 1}
        & \mathbf{0}_{K\times 1}
        \\
        \delta_t^{k} \left( \displaystyle Bo \bfjfm{I} - \textstyle Sr\, \textstyle Dr \bfjfm{\Delta_H}\right)
        & a^k \bfjfm{I} - \frac{ \textstyle 2 \delta^k_t}{ \textstyle \Rey} \bfjfm{\Delta_H}
        & \delta_t^{k}Dr \bfjfm{I}
        & \mathbf{0}_{K\times 1}
        & \mathbf{0}_{K\times 1}
        \\
        \mathbf{0}_{1\times K}
        & \mathbf{0}_{1\times K}
        & \frac{\textstyle 3\delta^k_t}{\textstyle 4\pi} \bfjfm{S}(q)
        & a^k
        & 0
        \\
        \mathbf{0}_{1\times K}
        & \mathbf{0}_{1\times K}
        & \mathbf{0}_{1\times K}
        & -\delta^k_t
        & a^k
    \end{array}
    \right],
 \end{equation}

 \begin{equation}
    \label{eqn:W_independent_vector_Def}
     W^{k+1} = 
     \left[
     \begin{array}{ccccc}
          \eta^{k+1} & \phi^{k+1} & p^{k+1} & v^{k+1} & h^{k+1}
     \end{array}
     \right]^{\top},
 \end{equation}
\begin{equation}
     \label{eqn:bath_matrix_form_2}
     \bfjfm{R}(\delta_t^k) 
     =
     \left[
    \begin{array}{ccc}
        a^k\bfjfm{I} 
        & -\delta_t^k\bfjfm{I}
        & \bfjfm{0}
        \\
        \delta^k_t\bfjfm{P}
        & 
        a^k \bfjfm{I} + \delta^k_t \bfjfm{T}
        & 
        \delta^k_t \bfjfm{V}
    \end{array}
    \right],
 \end{equation}
 with
 \begin{equation}
     P_{i, j} = (j+1)(j+3)j\delta_{ij}
             , \quad \text{ for } i, j = 1 \dots M
 \end{equation}

  \begin{equation}
     T_{i, j} =  (2j+3)j\delta_{ij}
             , \quad \text{ for } i, j = 1 \dots M
 \end{equation}

\begin{equation}
     V_{i, j} =  (j+1)\delta_{ij}
             , \quad \text{ for } i, j = 1 \dots M
 \end{equation}
 
  \begin{equation}
  \label{eqn:Y_definition}
     Y^{k+1} = 
     \left[
     \begin{array}{ccc}
          A^{k+1} & U^{k+1} & B^{k+1}
     \end{array}
     \right]^{\top},
 \end{equation}

 \begin{equation}
     F = 
     \left[
     \begin{array}{c}
          - b^k \eta^{k} - c^k \eta^{k-1} \\
          -b^k \phi^{k} - c^k \phi^{k-1} \\
          -b^k v^k - c^k v^{k-1} \\
          -b^k h^k - c^k h^{k-1}
     \end{array}
     \right],
     \quad
     G =
     \left[
     \begin{array}{c}
          - b^k A^{k} - c^k A^{k-1} \\
          -b^k U^{k} - c^k U^{k-1} \\
     \end{array}
     \right].
 \end{equation}

where $\bfjfm{I}$ and $\bfjfm{0}$ are the identity and zero matrix of size $M$, respectively. We note that $F$ and $G$ depend on information from previous times only. The linear operators in equation (\ref{eqn:matrix_form_uncoupled}) represent the \emph{linear} bath and drop dynamics, respectively. They are coupled via equations (\ref{eqn:summary_discretised_restrictions}), which describe the free surface height compatibility with respect to the drop deformation during contact. This compatibility introduces nonlinearities in the system. 

We note that $\mathbf{R}$ is a rectangular matrix of size $2M \times 3M$, and $\mathbf{Q}$ is of size $(2K+2) \times (3K+2)$, corresponding to $2K+2$ equations and $3K+2$ unknowns. Further manipulations are thus needed to make them square matrices. 

The same procedure described by \cite{GaleanoRiosEtAl2017} to reduce the number of unknowns from $3K+2$ to $2K+2$ for the case of a rigid sphere is applied to this bath-drop matrix system. As such, we apply relations (\ref{eqn:zero_pressure_discretised}) and (\ref{eqn:discrete_contact_condition}) to reduce the effective number of unknowns.

This substitution makes the right-hand side of the first equation $A^{k+1}$-dependent. We obtain a modified matrix system,
\begin{equation}\label{eqn:matrix_form_reduced_1}
    \mathbf{Q}'(\delta_t^k, q) W'^{k+1} = F'(A^{k+1}, q) = F'. 
\end{equation}

Similarly, we recast the other matricial system by incorporating $B^{k+1}$ into the left-hand side of equation (\ref{eqn:matrix_form_uncoupled}), therefore having the following formulation for the reduced system:

\begin{equation}
\label{eqn:matrix_form_reduced_2}
\bfjfm{R}' (\delta_t^k) Y'^{k+1} = G'(\delta_t^k, B^{k+1});
\end{equation}
where

\begin{equation}
     \label{eqn:bath_matrix_form_3}
     \bfjfm{R}'(\delta_t^k) 
     =
     \left[
    \begin{array}{cc}
        a^k\bfjfm{I} 
        & -\delta_t^k\bfjfm{I}
        \\
        \delta^k_t\bfjfm{X}
        & a^k \bfjfm{I} 
    \end{array}
    \right],
 \end{equation}

\begin{equation}
     X_{i, j} =  (i+1)(i+3)i \ \delta_{ij}
             , \quad \text{ for } i, j = 1 \dots M,
 \end{equation}
 
  \begin{equation}
     Y'^{k+1} = 
     \left[
     \begin{array}{cc}
          A^{k+1} & U^{k+1}
     \end{array}
     \right]^{\top},
 \end{equation}

 \begin{equation}
     G'(\delta_t^k, B^{k+1}) =
     \left[
     \begin{array}{c}
          - b^k A^{k} - c^k A^{k-1} \\
          -b^k U^{k} - c^k U^{k-1}  - 
          \left(\begin{array}{c}
               2\delta_t^k B^{k+1}_2  \\
               3\delta_t^k B^{k+1}_3  \\
               \vdots \\
               (M+1)\delta_t^k B^{k+1}_{M+1}  \\
          \end{array}
          \right) \\
     \end{array}
     \right].
 \end{equation}

Equations (\ref{eqn:matrix_form_reduced_1}) and (\ref{eqn:matrix_form_reduced_2}) are therefore analogous to equations (3.19) and (3.6) from \cite{AgueroEtAl2022} and \cite{GaleanoRiosEtAl2017}, to render them square matrices. 

Further technical details regarding the computational implementation are given in Appendix \ref{sec:comp_implementation}. To simplify notation, deformation variables $B_l$ and $A_l$  without the $k$ superscript in subsequent sections will refer to the numerical solutions, unless stated otherwise.

\section{Computational implementation} \label{sec:comp_implementation}

\subsection{An iteration on pressure}
\label{sec:iteration_pressure}
The system of equations (\ref{eqn:matrix_form_reduced_1}), and (\ref{eqn:matrix_form_reduced_2}), has to be solved together with the nonlinear set of equations given by (\ref{eqn:summary_discretised_restrictions}). We address the interdependence of $A^{k+1}$ in both (\ref{eqn:matrix_form_reduced_1}) and (\ref{eqn:matrix_form_reduced_2}) by using the iterative procedure outlined in Algorithm \ref{alg:IterationOnPressure} to compute $B^{k+1}$.

\begin{algorithm} 
    \DontPrintSemicolon
    \SetKwInOut{Input}{Input}
    \SetKwInOut{Output}{Output}
    \Input{$\delta^k_t$, q, $\eta^k$, $\phi^k$, $p^k$, $h^k$, $v^k$, $B^k$, $A^k$, $U^k$, $\eta^{k-1}$, $\phi^{k-1}$, $p^{k-1}$, $h^{k-1}$, $v^{k-1}$, $B^{k-1}$, $A^{k-1}$ and $U^{k-1}$.}
    \Begin{
    set $B^{k+1, 0} \longleftarrow 0$\;
    set $B^{k+1, 1} \longleftarrow B^{k}$\;
    int j = 1 \;
    set $\varepsilon > 0$ \tcp{tolerance to exit outer loop}
    \tcp{The metric below could be the $\ell_2$ norm}
    \While{$d(B^{k+1, j}, B^{k+1, j-1}) \geq \varepsilon$}{
        \tcp{Solve the system below with known pressure}
        $Y'^{k+1} = (\bfjfm{R}')^{-1} G'(B^{k+1, j})$ \ \ \  \tcp{$\bfjfm{R}'$ is a $2M \times 2M$ system}
        \tcp{Solve the $(2K+2) \times (2K+2)$ system below}
        $W'^{k+1} = (\bfjfm{Q}')^{-1} F'(A^{k+1})$ \tcp{$A^{k+1}$ extracted from $Y'^{k+1}$}
        $p'_d = W'^{k+1}((2K+1-q):2K)$ \tcp{Extract pressure from solution}
        \For{\texttt{$l = 2, \dots, M+1$}}{
            $B^{k+1, j+1}_l = \bfjfm{L}(l) p'_d$ \tcp{Project pressure amplitudes}
        } 
        $j =j+1$\;
    } 
    } 
    \Output{$\eta^{k+1}$, $\phi^{k+1}$, $p^{k+1}$, $h^{k+1}$, $v^{k+1}$ (extracted from $W'^{k+1}$); $A^{k+1}$ and $U^{k+1}$ (extracted from $Y'^{k+1}$), and $B^{k+1} = B^{k+1, j}$.}
    
    \caption{Pseudocode \emph{solve\_system()} to find the coupled solution to (\ref{eqn:matrix_form_uncoupled}) by iterating on $B^{k+1, j}$}
    \label{alg:IterationOnPressure}
\end{algorithm}

Starting from an initial guess $B^{k+1,1}$, we iteratively compute $B^{k+1,j}$ for $j\ge1$, stopping when convergence yields the pressure coefficient vector $B^{k+1}$, which by construction satisfies both matrix systems at once.

The above-mentioned manipulations transform the problem of finding the solution of the next time step to finding the limit of $B^{k+1, j}$ in the following iterative subroutine:

\begin{align}
    B^{k+1,1} &= B^k, \\
     Y^{k+1, j} &= 
     \left(\bfjfm{R}'(\delta_t^k, q)\right)^{-1}G'(\delta_t^k, B^{k+1,j}), \\
     A^{k+1, j} &= A^{k+1, j}\left(Y^{k+1, j}\right),
     \quad \quad &\text{via (\ref{eqn:Y_definition})},\\
     W'^{k+1, j} &= 
     \left(\bfjfm{Q}'(\delta_t^k, q)\right)^{-1}F'(\delta_t^k, A^{k+1, j}), \\
    p^{k+1,j} &=  p^{k+1,j}(W'^{k+1}), 
     \quad \quad &\text{via (\ref{eqn:W_independent_vector_Def})},\\
    B^{k+1,j+1}_l &= \bfjfm{L}(l) p^{k+1,j}.
    \label{eqn:discretised_inner_loop}
\end{align}

The set of equations  \ref{eqn:discretised_inner_loop} is prescribed under the assumption that the contact area for that time step $t^k$ contains exactly $q$ contact points in the discretised mesh, which is itself an unknown to be found using another iterative method described later. 

\subsection{An iteration on geometry}

The contact radius is parametrized by \(q\) and remains to be determined. We note that from the system of restrictions (\ref{eqn:summary_discretised_restrictions}), only (\ref{eqn:discrete_condition_no_overlay_drop}) and (\ref{eqn:summary_km_tangent_condition}) have not been verified in the previous discussion. These two conditions will determine which value for $q$ is assigned to $r^{k+1}_c/\delta_r$.

\newcommand{\preq}[1]{\prescript{}{q}{#1}}
By following the procedure described in section \ref{sec:iteration_pressure}, each triplet $(k, \delta_t^k, q)$ determines a set of candidate solutions for the next time step, $\preq{\eta^{k+1}}$, $\preq{\phi^{k+1}}$, $\preq{p^{k+1}}$, $\preq{h^{k+1}}$, $\preq{v^{k+1}}$, $\preq{r^{k+1}_{c}}$, $\preq{B^{k+1}}$, $\preq{A^{k+1}}$ and $\preq{U^{k+1}}$. 

We define the error of this potential solution by the functional

\begin{equation}
    \label{eqn:error_tangent_definition}
    e(q, \delta_t^k) =
    \begin{cases}
        \infty & \text{if (\ref{eqn:discrete_condition_no_overlay_drop}) is violated.} \\ \\
             \left|\frac{\preq{\eta^{k+1}_{q+1}}-\preq{\eta^{k+1}_{q}}}{\delta_r} - \partial_r \preq{z_d}((q-1/2)\delta_r)\right| & \text{otherwise}        
    \end{cases}
\end{equation}

 After convergence of the pressure iteration $B^{k+1,j}\to B^{k+1}$, condition (\ref{eqn:error_tangent_definition}) is used to eliminate any solution where the drop and bath surfaces intersect. For the remaining candidates, the residual in (\ref{eqn:summary_km_tangent_condition}) provides a quantitative measure of the error associated with the parameters $(\delta_t^k, q)$. An exact partial differential equation solution would drive this residual to zero by satisfying (\ref{eqn:summary_km_tangent_condition}) exactly. We then select the number of contact points at $t^{k+1}$ that minimizes this residual to define the pressed area,
\begin{equation}
    r^{k+1}_c := \delta_r \cdot \text{argmin}_q \ e(q, \delta_t^k).
\end{equation}

\begin{algorithm} 
    \DontPrintSemicolon
    \SetKwInOut{Input}{Input}
    \SetKwInOut{Output}{Output}
    \Input{$\eta^1$, $\phi^1$, $p^1$, $h^1$, $v^1$, $B^1$, $A^1$ and $U^1$}
    \Begin{
    int k = 0; float $t^k$ = 0 \;
    $\mathbf{t}$ = $linspace(0, t_{final}, 5000)$ \tcp{Initial time vector}
    \While{$t^k < t_{final}$}{
        
        $\delta_t^{k} = \mathbf{t}(k+1) - \mathbf{t}(k) $ \;
            \For{$q \gets r^k_c/\delta_r - 2$ \KwTo $r^k_c/\delta_r + 2$ }{
                $\preq{\eta}^{k+1}$, $\preq{\phi}^{k+1}$, $\preq{p}^{k+1}_d$, $\preq{h}^{k+1}$, $\preq{v}^{k+1}$, $\preq{B}^{k+1}$, $\preq{A}^{k+1}$, $\preq{U}^{k+1}$ \\
                $\quad \quad = 
                \text{Algorithm\_1(} \delta^k_t, q, \eta^k, \phi^k, p^k, h^k, v^k, B^k, A^k, U^k$ \\
                $\quad \quad \quad \quad \eta^{k-1}$, $\phi^{k-1}$, $p^{k-1}$, $h^{k-1}$, $v^{k-1}$, $B^{k-1}$, $A^{k-1}$, $U^{k-1}$
                )\;
                       
            \eIf{$\preq{\eta^{k+1}_i} > \preq{h^{k+1}} + \preq{z^{k+1}_d(\delta_r \cdot i)}$ } 
            {
                $e(q, \delta_t^k) = \infty$ \tcp{Following (\ref{eqn:no_intersection_condition})}
            }{
                $e(q, \delta_t^k) = \left|\frac{\preq{\eta^{k+1}_{q+1}}-\preq{\eta^{k+1}_{q}}}{\delta_r} - \partial_r \preq{z_d}((q+1/2)\delta_r)\right|$;
            }
            } 

            $r^{k+1}_c \longleftarrow \delta_r \cdot \text{argmin}_q \ e(q, \delta_t^k)$ \;
        \eIf{$|r^{k+1}_c - r^k_c| > \delta_r$}{
            $\mathbf{t} = insert(\mathbf{t}, k+1, (\mathbf{t}(k) + \mathbf{t}(k+1))/2$);
            \tcp{Insert intermediate time step}
        }{
            $k = k + 1$; \tcp{Accept solution and advance time step}
        }
       
    } 
    
    } 
    \Output{
    $\eta^k$, $\phi^k$, $p^k$, $h^k$, $v^k$, $B^k$, $A^k$, $U^k$, $\forall k\geq 1$, and $\mathbf{t}$
    }
    
    \caption{Pseudocode to solve the discrete formulation at the next time step}\label{alg:IterationOnGeometry}
\end{algorithm}

\subsection{Numerical implementation considerations}

To guarantee a physically continuous evolution of the pressed surface \(S(t)\), we enforce a smooth update of the discretised contact radius via
\[
\bigl\lvert r^{k+1}_c - r^{k}_c \bigr\rvert \le \delta_r.
\]
Whenever this criterion is violated, we halve the time step (\(\delta_t^k \leftarrow \delta_t^k/2\)) and discard all candidate solutions computed with the larger step.  The search then restarts under the refined temporal resolution, ensuring that each update of the contact area proceeds without jumps.

The numerical implementation further ensures that surface waves do not reach the boundaries of the computational domain, preserving the validity of the imposed assumptions. Additionally, condition \(|\nabla \eta| < 1\) is verified for all solutions at all times, ensuring the free surface slope remains within the linear limit consistent with the linearised curvature approximation.

As already described, time derivatives are discretised with a second-order backward difference formula (BDF2).  To bootstrap BDF2, the first increment uses a first-order implicit Euler step, providing the two initial time levels needed for the higher-order scheme.

Integral operators  $\mathbf{S}(q)$ and $\bfjfm{L}(l)$ used in equations (\ref{eqn:summary_discretised_centerofmass_drop} )and (\ref{eqn:discrete_Legendre})  are implemented using a trapezoidal rule, as described in Appendix \ref{section:Appendix_L}. 

The simulations reported here use a spectral representation of the droplet with \(M=20\) modes (equation \ref{eqn:Legendre_decomposition_1}). This truncation level was chosen to ensure the stability of the inner pressure optimization loop, which becomes unstable for \(M \gtrsim 65\). To validate this choice, simulations were repeated with \(M=40\) and \(M=60\) modes, ensuring a relative change of at most 5\% in all relevant variables and yielding visually indistinguishable plots.

To sum up, the full algorithm is laid out in Algorithm \ref{alg:IterationOnGeometry}.  After initializing the solution and taking the first time step, the main loop generates multiple candidates (indexed by \(q\)), evaluates them against the tangent-error metric (\ref{eqn:error_tangent_definition}), and selects the optimal value of contact points to advance.  Any overly rapid numerical change in \(r^k_c\) triggers the aforementioned time-step reduction and recomputation.

\bibliographystyle{jfm}
\bibliography{Biblio_Impacts}
\end{document}